\documentclass[11pt,a4paper,longbibliography]{article}
\usepackage{jcappub} 

\usepackage{graphicx} 
\usepackage[utf8]{inputenc}
\usepackage{subcaption}
\usepackage{physics}
\usepackage{amsfonts,amssymb,amsmath}
\usepackage{mathtools} 
\usepackage[T1]{fontenc}
\usepackage{tikz}
\usetikzlibrary{positioning, shapes.geometric, arrows.meta, fit, calc, decorations.markings}
\usepackage{diagbox}
\usepackage{mathtools} 
\usepackage{multirow}
\usepackage{makecell} 
\usepackage{pdflscape}
\usepackage{fontawesome5} 

\usepackage{xcolor}

\newcommand{\hyperparams}{\Lambda}
\newcommand{\hyperpowerspectrum}{\Lambda_{\rm PS}}
\newcommand{\hyperGalaxies}{\Lambda_{\rm g}}
\newcommand{\hypermagnitudes}{\Lambda_{M}}

\newcommand{\hypercosmology}{\Lambda_{\rm c}}

\newcommand{\deffrom}{\coloneqq}
\newcommand{\p}{\mathbb{P}}

\newcommand{\Omegam}{\Omega_{{\rm m}}}

\newcommand{\hu}{\,{\rm km \,s^{-1} \, Mpc^{-1}}} %
\newcommand{\Mpc}{\,{\rm Mpc}} %

\newcommand{\densityDM}{\delta_{\rm DM}}

\newcommand{\countGalaxiesObs}{n_{\rm c}}

\newcommand{\countGalaxiesObspixel}{n_{{\rm c},I}}

\newcommand{\countGalaxiesTrue}{n_{\rm g}}
\newcommand{\countGalaxiesTruepixel}{n_{{\rm g},I}}

\newcommand{\Mthresh}{M_{\rm thr}}

\newcommand{\Deltastd}{\Delta_{\rm std}}
\newcommand{\Deltamedian}{\Delta_{\rm median}}

\newcommand{\gaussianfield}{F_{\rm G}}

\newcommand{\numpyro}{\texttt{numpyro}}
\newcommand{\jax}{\texttt{jax}}

\newcommand{\skypos}{\varphi_{\perp}}

\newcommand{\Data}{\mathcal{D}}

\newcommand{\catalogall}{\{\Data\}}
\newcommand{\nobsbin}{n_{{\rm c}, I}}
\newcommand{\nobsbinhat}{{n_{{\rm c},\hat I}}}

\newcommand{\ngtruebin}{n_{{\rm g},I}}

\newcommand{\rateAbs}{\mu_{\rm abs}}
\newcommand{\rateAbstilde}{\widetilde{\mu_{\rm abs}}}

\newcommand{\ratezhpM}{\mu(z,\skypos,M;\hyperparams)}
\newcommand{\ratezhp}{\mu(z,\skypos;\hyperparams)}
\newcommand{\ratezhpm}{\mu(z,\skypos,m;\hyperparams)}

\newcommand{\ratezhathpm}{\mu(\hat z,\skypos,m;\hyperparams)}
\newcommand{\ratezhpMofzandm}{\mu(z,\skypos,M(z,m;\hypercosmology);\hyperparams)}

\newcommand{\MPSamp}{M_{\text{PS,amp}}}
\newcommand{\Mpalph}{M_{p,\alpha}}
\newcommand{\Malphs}{M_{\alpha_s}}
\newcommand{\Mkz}{M_{k_0}}

\newcommand{\Dataprop}{\theta}
\newcommand{\catalog}{C}

\newcommand{\Mhnorm}{M_{h}}

\newcommand{\gladeplus}{\texttt{glade}+}

\newcommand{\mudet}{\mu_{\rm det}}
\newcommand{\sigmadet}{\sigma_{\rm det}}

\newcommand{\Xnorm}{X_{\rm norm}}

\newcommand{\mmean}{m_{\rm threshold}}
\newcommand{\mscale}{m_{\rm scale}}

\newcommand{\biasamp}{b_0}
\newcommand{\biasalpha}{\alpha_{\rm b}}
\newcommand{\biascut}{\beta_{\rm cut}}
\newcommand{\biasepsilon}{\epsilon_{\rm g}}

\newcommand{\kvec}{\vec{k}}
\newcommand{\keff}{||\kvec||}
\newcommand{\PDM}{\mathcal{P}_{\text{DM}}}
\newcommand{\PA}{P_{A,\text{DM}}}
\newcommand{\PnOne}{P_{n,1}}
\newcommand{\PnTwo}{P_{n,2}}
\newcommand{\PkEq}{P_{k,\text{eq}}}
\newcommand{\Pxi}{P_{\xi}}

\newcommand{\ratedet}{\mu({\det},\hat z, \skypos, m;\hyperparams)}

\newcommand{\Mthreshold}{M_{\rm thresh}}

\newcommand{\zsliceplot}{0.73}

\newcommand{\mightchange}[1]{#1}
\newcommand{\revvone}[1]{#1}
\usepackage[acronym, toc, nonumberlist]{glossaries}

\glsdisablehyper
\setacronymstyle{long-short}

\newacronym{ns}{NS}{neutron star}
\newacronym{bh}{BH}{black hole}
\newacronym{bbh}{BBH}{binary black hole}
\newacronym{bns}{BNS}{binary neutron star}
\newacronym{nsbh}{NSBH}{neutron star black hole}

\newacronym{eos}{EoS}{equation of state}
\newacronym{gw}{GW}{gravitational wave}
\newacronym{gr}{GR}{general relativity}
\newacronym{snr}{SNR}{signal-to-noise ratio}

\newacronym{lisa}{LISA}{Laser Interferometer Space Antenna }
\newacronym{ligo}{LIGO}{Laser Interferometer Gravitational wave Observatory}
\newacronym{kagra}{KAGRA}{KAmioka GRavitational wave detector}
\newacronym{eob}{EOB}{effective one-body}
\newacronym{em}{EM}{electromagnetic}
\newacronym{lcdm}{$\Lambda$CDM}{$\Lambda$ cold dark matter}
\newacronym{pl}{PL}{power law}
\newacronym{plg}{PLG}{power law and Gaussian}
\newacronym{kde}{KDE}{kernel density estimate}
\newacronym{de}{DE}{dark energy}
\newacronym{cdf}{CDF}{cumulative density function}

\newacronym{lvk}{LVK}{LIGO-Virgo-KAGRA}
\newacronym{ego}{EGO}{European gravitational observatory}
\newacronym{asd}{ASD}{amplitude spectral density}
\newacronym{psd}{PSD}{power spectral density}
\newacronym{mcmc}{MCMC}{Monte Carlo Markov chain}
\newacronym{hlv}{HLV}{Hanford Livingston Virgo}
\newacronym{pe}{PE}{parameter estimation}
\newacronym{cbc}{CBC}{compact binary coalescence}
\newacronym{aligo}{aLIGO}{advanced LIGO}
\newacronym{far}{FAR}{false alarm rate}

\newacronym{cl}{CL}{confidence level}

\newacronym{pn}{PN}{post-Newtonian}
\newacronym{nr}{NR}{numerical relativity}
\newacronym{ppisn}{PPISN}{pulsation pair-instability supernova}
\newacronym{pisn}{PISN}{pair instability-supernova}
\newacronym{et}{ET}{Einstein Telescope}
\newacronym{ce}{CE}{Cosmic Explorer}

\newacronym{cmb}{CMB}{cosmic microwave background}
\newacronym{lss}{LSS}{large scale structure}
\newacronym{isco}{ISCO}{innermost stable orbit}

\newacronym{oi}{Oi}{observation run $i$}
\newacronym{gwtci}{GWTC-i}{gravitational wave transient catalog $i$}

\newacronym{2g}{2G}{second generation}
\newacronym{3g}{3G}{third generation}

\newacronym{bao}{BAO}{baryonic acoustic oscillation}

\newacronym{wkb}{WKB}{Wentzel–Kramers–Brillouin}

\newacronym{dpg}{DPG}{Dvali-Gabadadze-Porrati}
\newacronym{dhost}{DHOST}{degenerate higher-order scalar-tensor}
\newacronym{mg}{MG}{modified gravity}
\newacronym{des}{DES}{dark energy survey}

\newacronym{tt}{TT}{transverse-traceless}

\newacronym{sgwb}{SGWB}{stochastic gravitational wave background}
\newacronym{dgrb}{DGRB}{diffuse $\gamma$-ray background}
\newacronym{gut}{GUT}{grand unified theory}

\newacronym{ng}{NG}{Nambu-Goto}
\newacronym{gbr}{GBR}{gravitational backreaction}

\newacronym{nf}{NF}{normalizing flow}
\newacronym{ml}{ML}{machine learning}
\newacronym{lfi}{LFI}{likelihood-free inference}
\newacronym{nn}{NN}{neural network}
\newacronym{dingo}{DINGO}{deep inference for gravitational wave observations}
\newacronym{gpu}{GPU}{graphics processing unit}
\newacronym{hba}{HBA}{hierarchical Bayesian analysis}

\newacronym{kl}{KL}{Kullback-Leibler}
\newacronym{js}{JS}{Jensen-Shannon}
\newacronym{ks}{KS}{Kolmogorov–Smirnov}

\newacronym{smbh}{SMBH}{supermassive black hole}
\newacronym{agn}{AGN}{active galactic nuclei}
\newacronym{spa}{SPA}{stationary phase approximation}

\newacronym{pta}{PTA}{pulsar timing array}

\newacronym{npe}{NPE}{neural posterior estimation}

\newacronym{dm}{DM}{dark matter}

\newacronym{grf}{GRF}{Gaussian random field}

\newacronym{desi}{DESI}{Dark Energy Spectroscopic Instrument}

\title{Cosmic Cartography II: completing galaxy catalogs for gravitational-wave cosmology}
\author{}
\date{September 2024}

\author{Konstantin Leyde$^a$,}
\author{Tessa Baker$^a$,}
\author{Wolfgang Enzi$^a$}
\affiliation[a]{Institute of Cosmology and Gravitation, University of Portsmouth, \\
Burnaby Road, Portsmouth PO1 3FX, United Kingdom}
\emailAdd{konstantin.leyde@port.ac.uk}
\emailAdd{tessa.baker@port.ac.uk}
\emailAdd{wolfgang.enzi@port.ac.uk}

\begin{abstract}{
The dark siren method exploits the complementarity between gravitational-wave binary coalescence signals and galaxy catalogs originating from the same regions of space.
However, all galaxy catalogs are incomplete, i.e.~they only include a subset of all galaxies, typically being biased towards the bright end of the luminosity distribution. 
This sub-selection systematically affects the dark siren inference of the Hubble constant $H_0$, so a completeness relation has to be introduced that accounts for the missing objects. 
In the literature it is standard to assume that the missing galaxies are uniformly distributed across the sky and that the galaxy magnitude distribution is known.
In this work we develop a novel method which improves upon these assumptions and reconstructs the underlying true galaxy field, respecting the spatial correlation of galaxies on large scales. In our method the true magnitude distribution of galaxies is inferred alongside the spatial galaxy distribution. Our method results in an improved three-dimensional prior in redshift and sky position for the host galaxy of a GW event, which is expected to make the resulting $H_0$ posterior more robust. Building on our previous work, we make a number of improvements, and validate our method on simulated data based on the Millennium simulation. 
The inference results can be reproduced through our publicly available code base \href{https://github.com/KonstantinLeyde/light}{\faGithub \;light}.
}
\end{abstract}

\begin{document}
\maketitle

\section{Introduction}

\subsection{Current state of the field}

\Glspl{gw} provide an independent means of probing the expansion history of the Universe. This is particularly interesting given the latest discrepancies in the cosmological parameters, both in the current expansion rate of the Universe, $H_0$ \cite{Planck:2018vyg, Riess:2021jrx}, and the evolving equation of state of dark energy \cite{DESI:2024mwx, DESI:2025zgx}.
The extent to which GWs can help resolve these growing tensions is now a central question in the field \cite{LIGOScientific:2018gmd, Borhanian:2020vyr, Borghi:2023opd, Hanselman:2024hqy, Wang:2024zfv, Chen:2024gdn, Tagliazucchi:2025ofb}. 
\Gls{lvk} \Gls{gw} data \cite{det1-aligo2015,det2-aLIGO:2020wna,det3-Tse:2019wcy,det4-VIRGO:2014yos,det5-Virgo:2019juy} have been used to measure the Hubble constant through the multi-messenger event GW170817 \cite{LIGOScientific:2017vwq, LIGOScientific:2017adf}, and also through the analysis of the \gls{gw} transient catalog GWTC-3 \cite{KAGRA:2021vkt}, resulting in $H_0=68^{+12}_{-8}\,\hu$ \cite{LIGOScientific:2021aug}.
For the Hubble constant measurement it is a large advantage to have precise redshift information, that can either originate from a direct measurement of the host galaxy redshift \cite{LIGOScientific:2017vwq, LIGOScientific:2017adf, Howlett:2019mdh, Chen:2020dyt, Chen:2023dgw, Palmese:2023beh, Mancarella:2024qle, Salvarese:2024jpq} (i.e.~the bright siren method), from a statistical measurement using a galaxy catalog \cite{Schutz:1986gp, DelPozzo:2011vcw, LIGOScientific:2018gmd, Gray:2019ksv, DES:2020nay, Gray:2021sew, Palmese:2021mjm, Finke:2021aom, Turski:2023lxq, Mastrogiovanni:2023emh, DESI:2023fij, Alfradique:2023giv, Gray:2023wgj, Gair:2022zsa, Perna:2024lod, Beirnaert:2025wcx, Naveed:2025kgk} (i.e.~the dark siren method), or the redshifted source-frame mass distribution \cite{Taylor:2011fs, Taylor:2012db, Farr:2019twy, Mastrogiovanni:2021wsd, Mancarella:2021ecn, Leyde:2022orh, Ezquiaga:2022zkx, Pierra:2023deu, Leyde:2023iof, Bom:2024afj, Farah:2024xub, MaganaHernandez:2024uty, Mali:2024wpq, Agarwal:2024hld, Li:2024rmi, Tong:2025xvd} (i.e.~the spectral siren method).

Dark siren codes for $H_0$ inference \cite{Mastrogiovanni:2023emh, Gray:2023wgj, Borghi:2023opd} now combine and leverage both galaxy and source-frame mass distributions into cosmological measurements, although the galaxy catalog information currently provides little constraining power \cite{LIGOScientific:2019zcs, LIGOScientific:2021aug, Mastrogiovanni:2023emh, Gray:2023wgj}.
This is usually attributed to (\textit{i}) the incompleteness of the available galaxy catalogs (i.e.~faint galaxies are only seen in the near-by Universe) as well as (\textit{ii}) most GW signals are not well-localized in the sky or luminosity distance. 
In case of (\textit{i}) little redshift information for the GW is provided, and hence, the constraint on $H_0$ is weak. 
Even if the catalog is complete, a large GW sky position (point (\textit{ii})) leads to averages of the galaxy redshift prior over large volumes. Since the Universe is homogeneous on large scales, this case is asymptotically equivalent (in the limit of no sky position information) to using solely the marginalized redshift distribution of galaxies.
Thus, poorly localized GW events yield weaker constraints on cosmological parameters when compared to well-localized events.

The latest LVK cosmological analyses \cite{LIGOScientific:2021aug, LIGOScientific:2025jau} leverage GW data in combination with the set of galaxy catalogs called \gladeplus{} \cite{Dalya:2018cnd, Dalya:2021ewn}. This set of catalogs is essentially empty in the K-band at $z\sim 0.3$. 
Since many GW signals occur at higher equivalent redshifts (equivalent assuming a reference cosmology to convert the GW luminosity distance to a redshift), the catalog incompleteness poses a bottleneck. 
To formulate a completeness correction (i.e.~what fraction of galaxies is observed for a given redshift), a galaxy magnitude distribution is required, the parameters of which are fixed in all standard codes, although this distribution is not known perfectly. 
Additionally, the aforementioned codes \cite{Mastrogiovanni:2023emh, Gray:2023wgj, Borghi:2023opd} implement the completion correction assuming a uniform-in-comoving-distance distribution for the \gls{gw} host. 
While it is acknowledged that the missing galaxies are not uniformly scattered, \revvone{limited} progress has been made to complete galaxy catalogs disposing of this unphysical assumption. \revvone{For methodological development for a full galaxy catalog-GW analysis see \cite{2019arXiv191112337L}, and refer to \cite{Oguri:2016dgk, Bera:2020jhx, Mukherjee:2022afz,  Ghosh:2023ksl, Begnoni:2024tvj, Afroz:2024joi, Ghosh:2024cwc, Ferri:2024amc, Pedrotti:2025tfg} for a vivid recent interest treating the cross-correlation of GWs with galaxies (both for current and future detector networks, forecasting $H_0$ measurements at percent-level precision). 
Finally the work of \cite{Finke:2021aom} has introduced a phenomenological prescription to account for spatial correlation between galaxies, while a completion technique based on the variance of the galaxy number counts in three-dimensional volumes of similar completion has been put forward in \cite{Dalang:2023ehp, Dalang:2024gfk}. }

Our works solves the aforementioned problems. We address the incompleteness, by reconstructing the full galaxy field given a magnitude-limited catalog, under the hypothesis that the galaxy field respects a given two-point statistic (accounting for spatial correlation).
We also marginalize over the galaxy magnitude distribution; it is jointly inferred along with the galaxy field. 
This leads to a more informative galaxy redshift-sky position prior for GW events that implies a more informative dark siren $H_0$ measurement, schematically illustrated in Sec.~\ref{subsec: motivation}.
While this work demonstrates the reconstruction on simulated data it prepares the method for its application to \gladeplus{}. 



\subsection{Representative toy model}

\label{subsec: motivation}

We present here a toy model case to demonstrate the impact of ``completing the galaxy catalog'' on $H_0$ constraints using \gls{gw} data. This model is for illustration and is solely used in this section. 

\begin{figure}[!ht]
    \centering
    \includegraphics[width=\linewidth]{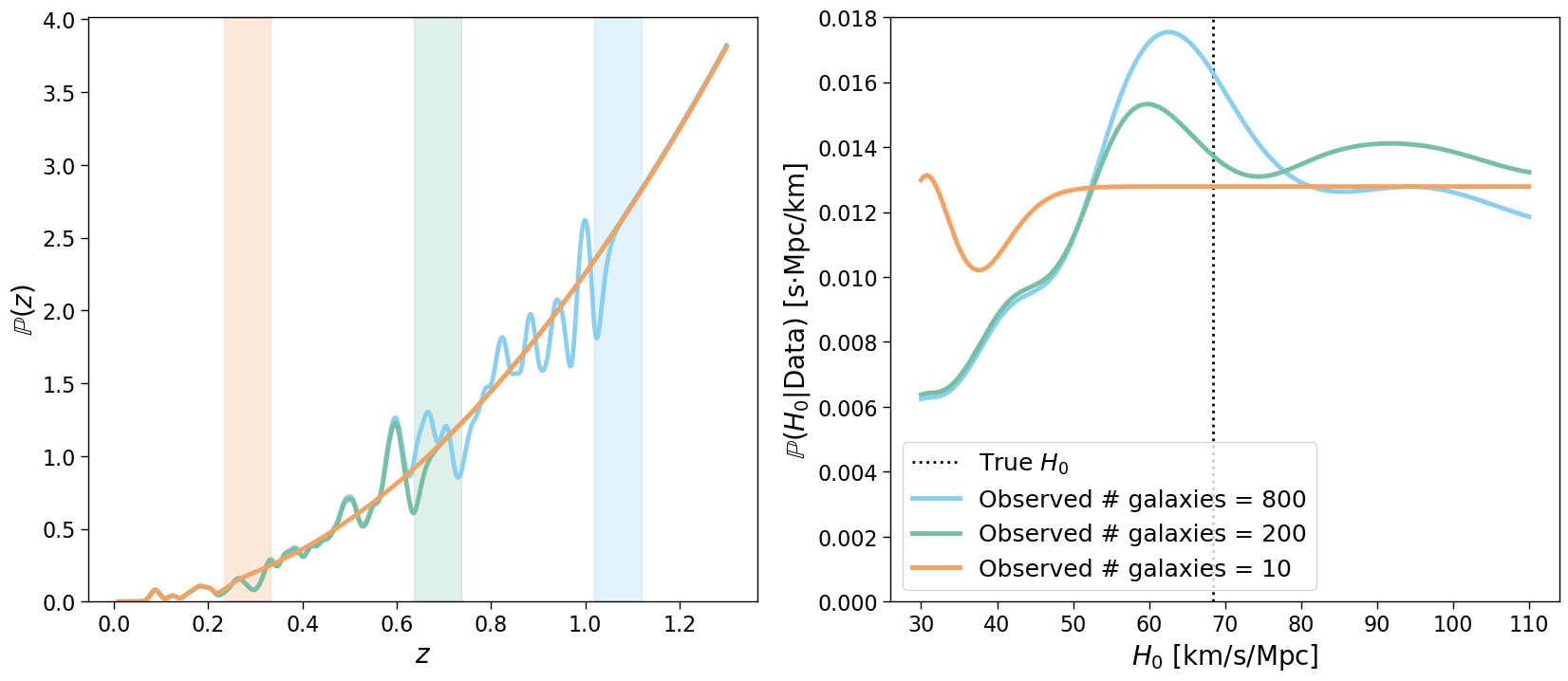}
    \caption{Toy model: schematic impact of the galaxy catalog completeness.
    \textit{Left panel}: The redshift prior for one fiducial sky pixel, where different colors indicate different completeness. 
    The catalogs transition to the homogeneous distribution at different redshifts: while the catalog of 10~galaxies (\mightchange{orange}) reverts to the homogeneous distribution already at $z\sim0.2$, the catalog of 800 galaxies (\mightchange{blue}) only transitions at $z\sim 1$. 
    This has important consequences on the $H_0$ posterior as the right panel illustrates: while the most incomplete catalog provides an almost uninformative $H_0$ posterior (\revvone{orange}), the more complete catalogs with \mightchange{200} and \mightchange{800} galaxies (\revvone{green and blue}), respectively, lead to a more informative $H_0$ measurement.
    }
    \label{fig:completeness impact schematic}
\end{figure}

For simplicity, we linearize the relationship between redshift and luminosity distance, i.e.~$d_L = c\,z / H_0$.
All analyzed GW events are localized in the same sky pixel (i.e.~element of the sky) that contains $\mathcal{O}(10^3)$ galaxies (detected and non-detected).
To illustrate the impact of the galaxy completeness we then vary the number of observed galaxies included in the catalog. 
\revvone{We assume that the redshift uncertainty of each galaxy to follow a Gaussian distribution with standard deviation of \mightchange{$\sigma_{z}=0.01$}.}
From the observed galaxy catalog one can build the line-of-sight redshift prior, the sum of the individual galaxy redshift probability density distributions.\footnote{As pointed out in \cite{Gair:2022zsa} this treatment is approximate, since one should consider a hierarchical model for the galaxy distribution. Currently however, this treatment is standard in all inference codes. } 
Note that because the catalog is incomplete, the line-of-sight redshift prior has two contributions: the in-catalog part (informed by the observed galaxies) and out-of-catalog part (with the standard assumption that galaxies are distributed uniformly in comoving volume).
We construct the in-catalog part as a sum of Gaussian distributions with their respective redshift error that modulate the volumetric $z^2$ prior. The out-of-catalog part is simply modeled as a $z^2$ distribution. 
If we define the sigmoid 
\begin{equation}
    \label{eq: def sigmoid}
    {\rm sig}(x) \deffrom \frac{1}{1+\exp\left[-x\right]}\,,
\end{equation}
the function ${\rm sig}\left(\frac{z-z_{\rm thresh}}{0.01}\right)$, for different threshold values of $z_{\rm thresh}$ (depending on the observed catalog completeness) is used to interpolate between in-catalog and out-of-catalog parts. 
The resulting line-of-sight redshift priors of three such scenarios (with the redshift threshold value of \mightchange{$0.23$, $0.64$ and $1.02$, respectively}) are plotted on the left side of Fig.~\ref{fig:completeness impact schematic} -- the region at which each catalog becomes incomplete marked in the respective color.

We then simulate GW events with Gaussian uncertainties on the luminosity distance, with a relative uncertainty of \mightchange{$5\%$}. 
We also mimic a detection bias by only allowing GW events with a measured luminosity distance smaller than \mightchange{$d_{L} = 4000$\,Mpc}. Modeling a detection threshold in this way simplifies selection effects encountered for true data such as the inverse false alarm rates which gives detection horizons that vary with mass, inclination, etc.
The right side of Fig.~\ref{fig:completeness impact schematic} shows the resulting $H_0$ posterior distributions for \mightchange{four} detected GW events, analyzed with \mightchange{three} different galaxy catalogs.
The first catalog contains \mightchange{10} galaxies, resulting in a $H_0$ posterior closely following its prior (assumed to be uniform).
The second (third) catalog contains \mightchange{200} (\mightchange{800}) galaxies, resulting in a $H_0$ posterior that is more informative.
This illustrates the importance of a complete galaxy catalog: the resulting $H_0$ posteriors do not agree, i.e.~it matters whether we neglect the fluctuations of the galaxy density, even at high redshift. 
Hence, our work will enable one to measure $H_0$ more precisely from building an ``improved'' version of an incomplete catalog.

\begin{figure}[!ht]
     \centering
     \includegraphics[width=0.67\linewidth]{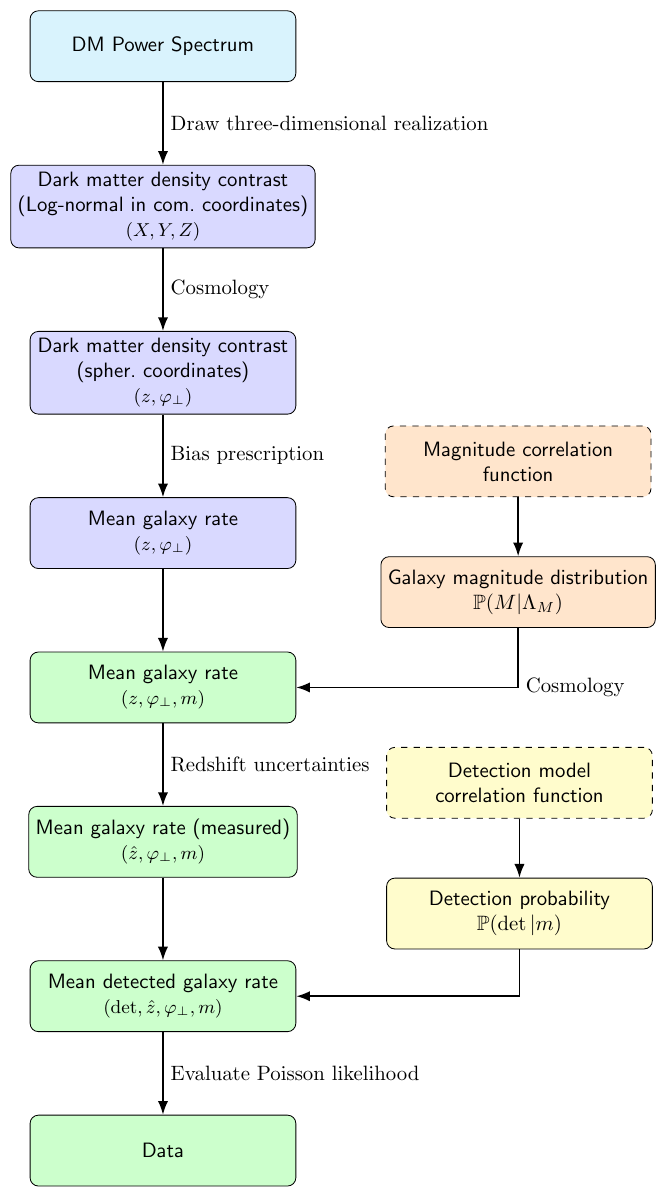}
     \caption{Schematic overview of the building blocks of the forward model.
     All quantities that appear in boxes with dashed lines are fixed prior to the analysis (e.g.~the parameters that govern the correlation structure of the magnitude distribution, cf.~App.~\ref{app:magnitude_distribution}). 
     Note however that all remaining variables (including the magnitude distribution) are inferred jointly along with the DM power spectrum and density contrast.
     }
     \label{fig:schematic overview}
 \end{figure}

\subsection{Novelty}
\label{subsec: novelty}

Let us here clearly draw up the differences between our previous method of \cite{Leyde:2024tov} and the present work. 
In our previous reconstruction we have estimated the galaxy number count per three-dimensional voxel. 
We had assumed a simplified setup, namely a cartesian box, a parametrized form for the galaxy magnitude distribution, and a local bias prescription linking \gls{dm} density contrast and mean galaxy rate. 
For the DM model itself, we assumed a log-normal field with a two-point statistic that was estimated but of a parametrized form. 
Here, we go significantly further and
\begin{itemize}
\setlength{\itemsep}{0pt}
    \item formulate the reconstruction in spherical coordinates,
    \item reconstruct the galaxy field in the additional magnitude dimension (i.e.~estimate the number of galaxies in each three-dimensional voxel, in each magnitude bin), using a flexible galaxy magnitude distribution that relies on a correlated Gaussian field, 
    \item include redshift uncertainties (compatible to spectroscopic-like surveys),
    \item jointly estimate the selection function of the survey (i.e.~what fraction of galaxies are observed for a given apparent magnitude). 
\end{itemize}
These improvements partially address challenges posed for the analysis of real data with our method, while we leave the development of a more sophisticated forward model for the dark matter distribution to future work.

\subsection{Paper structure}

The structure of this work is as follows. 
We begin by specifying our forward model in Sec.~\ref{sec: methodology}, describing how the log-normal field approximates the DM distribution, detailing the link between DM density contrast and galaxies, as well as defining the galaxy magnitude distribution. 
Sec.~\ref{sec: likelihood} puts all pieces together and formulates the likelihood. To validate the method, Sec.~\ref{sec: simulated data} details the construction of the observed catalog, based on the Millennium simulation.
In the results section (Sec.~\ref{sec: results}), we present the Bayesian reconstruction and test various model mis-specifications, such as Gaussian redshift uncertainties. 
The conclusions are presented in Sec.~\ref{sec: conclusions}.

We refer the busy reader to Fig.~\ref{fig:schematic overview} for an overview of the method, and to Sec.~\ref{sec:result vanilla} with Figures~\ref{fig: reconstruction phi-slice}, \ref{fig:mollweide slice in redshift} and \ref{fig:galaxy_counts_two_pixels} for the main results.

\begin{figure}[!ht]
     \centering
     \includegraphics[width=0.89\linewidth]{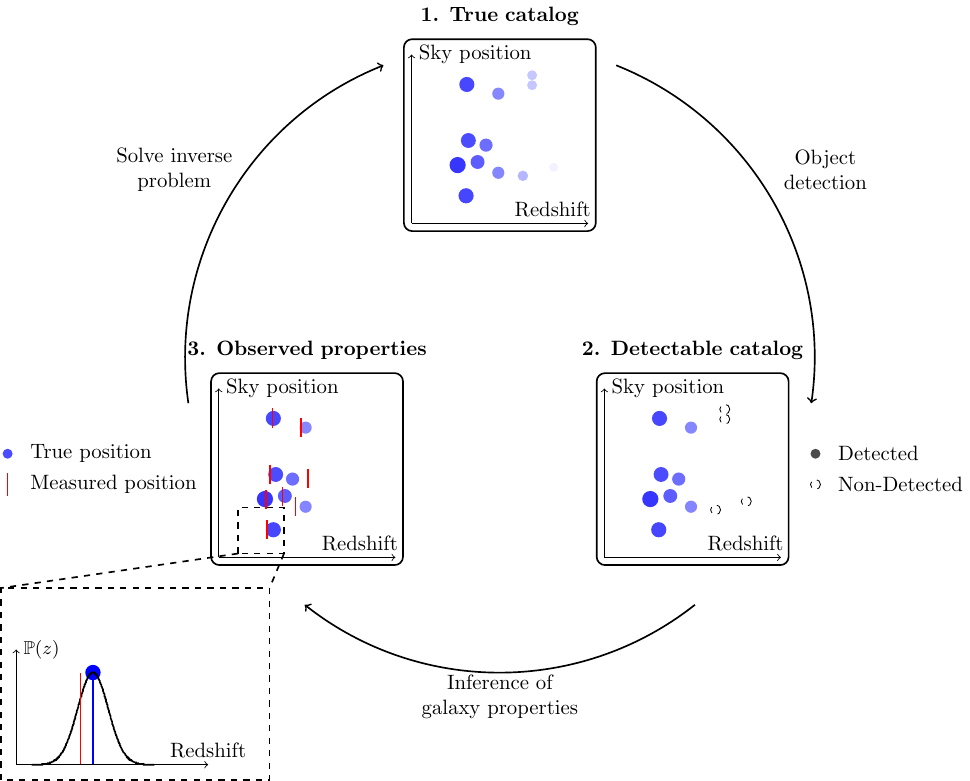}
     \caption{Schematic overview of the hierarchical inference problem.
     Starting from the true catalog, one identifies possible galaxies (step \textit{Object detection}). 
     We indicate the apparent magnitude (cf.~Eq.~\eqref{eq:def apparent magnitude}) of each galaxy with its \revvone{plotted color (blue), as well as its size}. Darker, \revvone{larger circles represent brighter galaxy in apparent magnitude}. 
     In a second step (step \textit{Inference \revvone{of} galaxy properties}) the parameters of each galaxy are inferred, independently for each object. 
     Finally, the inference of the individual galaxy properties is improved by considering a joint prior on the overall distribution of galaxies (step \textit{Solve inverse problem}).
     In this step, we start from the magnitude-limited catalog with the noisy galaxy properties and build possible configurations of the true catalog that are compatible with the observations.
     \revvone{Let us stress here that uncertainties in the galaxy sky position are neglected throughout this analysis. }
     }
     \label{fig:schematic overview hierarchical inference problem}
 \end{figure}

\section{Forward-modeling of galaxy catalog data}
\label{sec: methodology}

Ultimately, we want to improve the redshift-sky position prior for GW events. To this purpose, we need to construct a probability distribution in redshift, $z$ and sky position, $\skypos$, $\p(z,\skypos)$, given a tracer of \glspl{cbc}. 
Typically it is assumed that this probability increases with the counts of galaxies in each three-dimensional region of space (i.e.~voxel). 
However, to connect the galaxy distribution with the redshift-sky position PDF, we need to assume a probability for each galaxy to be a host of the GW source. 
As a proxy, it is assumed in all standard codes that this probability is proportional to the galaxy's luminosity in some chosen color band. 
Thus, it is not sufficient to construct the three-dimensional galaxy number count, but one also needs the galaxy number as a function of the absolute magnitude: the ultimate quantity we want to infer are the counts of galaxies per four-dimensional bin (redshift, sky position and absolute magnitude). 
To complicate the reconstruction further, the true catalog is stochastically related to the underlying \gls{dm} field (for instance, through galaxy formation), and observed data is noise-limited (the observed catalog is more likely to include brighter galaxies). 
Thus, there are many ``true'' catalogs and DM field realizations compatible with a given observed magnitude-limited catalog. 
Hence, we also want to infer the uncertainties of the reconstruction, i.e.~determine the posterior of the reconstructed galaxy number counts.  
In order to do so, we need a likelihood that connects the hyperparameters (that will be defined below) and the data.

The purpose of this section is thus to formulate the forward model that implicitly defines the likelihood of observing a catalog given 
\begin{itemize}
    \item the DM density field (in each three-dimensional voxel),
    \item the galaxy magnitude distribution,
    \item the bias parameters linking DM field with the galaxy field, and
    \item the detection probability of galaxies. 
\end{itemize}
We will refer to these parameters jointly as hyperparameters $\hyperparams$. Table~\ref{tab: overview variables} defines and describes the hyperparameters of each component above further. 

\begin{table}[]
    \centering
    \renewcommand{\arraystretch}{1.37}
    \begin{tabular}{cl}
        \hline
        \hline
        \textbf{Variable} & \textbf{Description} \\
        \hline
        \multicolumn{2}{c}{\textbf{Hyperparameters}}\\
        \hline
        $\hyperparams$ & The set of all hyperparameters, including \\
        $\hyperpowerspectrum$ & Power spectrum parameters \\
        $\hyperGalaxies$ & Bias parameters  \\
        $\hypermagnitudes$ & Parameters of the absolute magnitude distribution \\
        $\hypercosmology$ & Cosmological parameters \\
        $\densityDM$ & Dark matter density contrast \\
        \hline
        \multicolumn{2}{c}{\textbf{Cosmological Parameters}}\\
        \hline
        $H_0$ & Hubble constant \\
        $\Omegam$ & Matter density parameter \\
        \hline
        \multicolumn{2}{c}{\textbf{Large-Scale Structure}}\\
        \hline
        $\PDM$ & Dark matter power spectrum \\
        \hline
        \multicolumn{2}{c}{\textbf{Galaxy Catalogs}}\\
        \hline
        $\countGalaxiesTrue$ & True galaxy count \\
        $\countGalaxiesObs$ & Observed galaxy count \\
        $M$ & Absolute magnitude \\
        $m$ & Apparent magnitude \\
        $z$ & Redshift \\
        $\hat z$ & Maximum likelihood estimate of redshift \\
        $\skypos$ & Sky position \\
        $\Data$ & Data associated to each galaxy, i.e.~$\Data \deffrom \{\hat z,   \skypos, m \}$ \\
        $\ratezhathpm$ & Galaxy Poisson rate with $\hat z, \skypos$ and $m$\\
        \hline
        \multicolumn{2}{c}{\textbf{Bin notation}}\\
        \hline
        $\nobsbin$ & Observed number count in bin $I$ (true quantities) \\
        $\nobsbinhat$ & Observed number count in bin $\hat I$ (measured quantities) \\
        \hline
        \hline
    \end{tabular}
    \caption{Overview of the model variables.}
    \label{tab: overview variables}
\end{table}

Schematically, the likelihood can be written as $\p(\catalogall | \hyperparams)$, with $\catalogall\deffrom \{\Data_c\}_{c\in \catalog}$ the observed catalog where $c$ is a label running over all galaxies in the observed catalog $C$.
The raw catalog data comes in the form of fluxes in different bands, or spectra from which the galaxy properties (redshift, composition, sky position, star formation rates, etc.) are deduced. 
We will refer to these individual properties as $\Dataprop$. As such, $\Dataprop$ is a $\mathcal{O}(3-20)$-dimensional vector. 
One possible way to characterize each galaxy is through posterior samples that approximate $
    \Dataprop \sim \p(\Dataprop | \Data) \,,
$ the individual posterior of the galaxy properties.
The likelihood of observing the catalog data $\catalogall$ given the hyperparameters $\hyperparams$ can be computed through\footnote{In this expression we omit terms relating to the selection effect and that depend on the overall rate. This equation is for illustrating purposes only: this approach would be computationally prohibitive and will not be used in the later analysis. } 
\begin{equation}
    \p(\catalogall | \hyperparams)
     \propto 
     \int \prod_{c\in\catalog} \left[
        \dd \Dataprop_c \,\p(\Data_c | \Dataprop_c) 
        \right]
        \p(\{\theta_c\}_{c\in\catalog} | \hyperparams)
        \,,
\end{equation}
marginalizing the joint prior of the true catalog, $\p(\{\theta_c\}_{c\in\catalog} | \hyperparams)$, over the individual galaxy properties. 
While future computational acceleration might allow the evaluation of the above integral, it is not the aim of this work to marginalize over the properties of $\mathcal{O}(10^{6-8})$ galaxies. 

Therefore, we choose to represent the catalog data as the galaxies histogrammed over the sky, the redshift and the apparent magnitude. Since the redshift is uncertain, we bin each galaxy by its maximum-likelihood value in redshift.  
That is, our observable is a number count in each three-dimensional bin, which we denote as $\nobsbinhat$, where $\hat I$ is the index in redshift, sky position and magnitude, and the hat indicating that the galaxies are binned via their noisy maximum-likelihood properties.
In general, the maximum-likelihood values do not correspond to their true values, an effect accounted for in our modeling, as outlined below. 

We approximate our likelihood as the probability of obtaining the observed catalog's histogram binned by their maximum likelihood values, i.e.
\begin{equation}
\label{eq:approximation likelihood}
    \p(\catalogall | \hyperparams) \approx \p(\{\nobsbinhat\}|\hyperparams)\,.
\end{equation}
Two assumptions underlie the above approximation, namely (\textit{i}) the probability of $\p(\Data|\hyperparams)$ is constant in each bin and, (\textit{ii}) \revvone{the observed data of each galaxy can be fully described} by the bin in which it falls (here its maximum-likelihood value for its properties).
While the first assumption can be satisfied by using a fine grid over the galaxy properties, the second assumption is stronger. If we assume the $\Dataprop$ uncertainties to be known, this assumes the individual galaxy likelihood to be independent of other latent variables not included in the histogram, e.g.~the time of observation, the type of galaxy, etc.

In order to compute the full likelihood of Eq.~\eqref{eq:approximation likelihood} we will first need to write down the models for individual building blocks that separate this problem into simple steps (e.g.~the connection of the DM density with the galaxy Poisson rate, the absolute magnitude distribution, the detection process, etc.).
The formulation of the forward model then implies a likelihood that we use to sample from the posterior.  
Fig.~\ref{fig:schematic overview} summarizes the individual parts of this forward model.  

One can consider the problem we attempt to solve as a hierarchical inference problem where the individually obtained galaxy properties (which were independently inferred from the other galaxy data) are improved through the formulation of a joint prior of galaxy properties of position and magnitude, accounting for the three-dimensional spatial correlation and magnitude distribution. This is illustrated by Fig.~\ref{fig:schematic overview hierarchical inference problem}. 
\revvone{Note that the posterior estimate of the joint galaxy properties generally differs from the estimate for the galaxy properties separately (which both in themselves will depend on the prior knowledge). Since the former estimate takes into account population-level knowledge of the galaxy catalog it tends to reduce the individual galaxy parameter uncertainties, an effect that is also known as \textit{shrinkage} \cite{2019arXiv191112337L}.
}

\subsection{The galaxy Poisson rate}
\label{subsec: galaxy poisson rate}

The central modeling quantity for the reconstruction of the full galaxy field is the galaxy rate, denoted as $\ratezhpM$. 
It describes the expected number count for a given redshift $z$, sky position $\skypos$ and absolute magnitude $M$.
We refer to this as a ``rate'', because it will later appear in the likelihood as the Poisson rate.
Let us emphasize here that this is not a rate in the sense of ``counts per time'', but rather, ``counts per XYZ'', where XYZ stands for the three aforementioned variables of $z, \skypos$ and $M$. 

The rate $\ratezhpM$ falls into several contributions, namely 
\begin{equation}
\label{eq: def ratezhpM}
    \ratezhpM = \underbrace{\rateAbs}_{\text{Overall galaxy rate}\;\;} 
    \times
    \underbrace{\p(z,\skypos|\hyperparams)}_{\text{Spatial distribution}} 
    \times
    \underbrace{\p(M|\hyperparams)}_{\text{Magnitude distribution}}
    \,,
\end{equation}
where we will define \mightchange{second} term in Eq.~\eqref{eq:pdf galaxies} of Sec.~\ref{subsec: bias}; this distribution will depend both on the DM realization and the DM-galaxy bias. 
The above \mightchange{third} term will be detailed in Sec.~\ref{subsec: magnitude distribution} and in App.~\ref{app:magnitude_distribution}.
The above factorization is completely general, except for the assumption that the magnitude distribution to be independent of sky position, redshift and of the value of the DM density contrast in that three-dimensional voxel. 

Eq.~\ref{eq: def ratezhpM} encodes all relevant physics of the problem. For a choice of the hyperparameters (that is, a realization of the DM density contrast, a given magnitude distribution and a bias model), the rate of galaxies is a deterministic function. 
All probabilistic modeling for the galaxy number counts enters through the DM density contrast, the magnitude distribution, the bias and the randomness of the Poisson sampling. 
It is also clear from Eq.~\eqref{eq: def ratezhpM} that one can obtain marginalized expressions for the rate as $\ratezhp = \int \dd M \ratezhpM$, allowing one to formulate probability distributions over the spatial coordinates that are relevant as redshift-sky position priors for GW cosmology. To account for the \gls{cbc} host probability of the galaxies to vary with both absolute magnitude or redshift, one can include a weighting factor in the \mightchange{previous} integrand. 

We will first describe the spatial distribution of Eq.~\eqref{eq: def ratezhpM}, that is determined by the \gls{dm} density contrast and the DM-galaxy bias. 
To this purpose, the upcoming section summarizes the log-normal model that serves as a proxy for the \gls{dm} field, followed by an overview of the bias between DM the galaxy rate in Sec.~\ref{subsec: bias}.

\subsection{Dark matter model}
\label{subsec:DM model}

In order to reconstruct the full galaxy field, we choose to model the mean galaxy number in each three-dimensional voxel as a local function of the DM density contrast.\footnote{As discussed later in Sec.~\ref{subsec: bias}, this is an approximation, as the real galaxy formation process is a complex function of the halo size, galaxy history, environment etc. }
To produce the latter (spatially correlated) field in spherical coordinates, we now describe the log-normal field \cite{Coles:1991if} that models the DM density contrast. 
Since we produce DM realizations on a cartesian grid, but need to evaluate its density on a spherical grid, we detail our discretization scheme and how we interpolate from cartesian to spherical coordinates in App.~\ref{appendix:spherical_interpolation}. 

Following the methodology of \cite{Leyde:2024tov} based on \cite{Coles:1991if}, we generate a log-normal field with a given power spectrum on a cartesian grid. 
The power spectrum is computed using a phenomenological model that we introduce below in Eq.~\eqref{eq: def phenomenological power spectrum}, the parameters of which we fit jointly with the other remaining hyperparameters such as the DM density contrast. 
The log-normal field is generated by applying a transformation to a Gaussian random field, given by \cite{Coles:1991if, Agrawal:2017khv}
\begin{equation}
\label{eq:lognormal_transformation}
    \densityDM = \exp\left(\gaussianfield - \frac{\sigma_{\text{G}}^2}{2}\right) - 1 \,,
\end{equation}
where we denote the Gaussian random field as $\gaussianfield$, and $\sigma_{\text{G}}^2$ is the variance of the Gaussian random field (over many field realizations).
The Gaussian field is first constructed by sampling from a multivariate normal distribution with a covariance matrix derived from the desired power spectrum, exploiting an analytical connection between power spectrum of the log-normal field and its underlying Gaussian random field \cite{Coles:1991if}.
The log-normal transformation of Eq.~\eqref{eq:lognormal_transformation} ensures that the resulting field is bounded by $-1$ from below, as required for the DM density contrast. 
Within this formalism, we do not take into account the redshift dependence of the power spectrum.
\revvone{Upon the application to real data this will warrant additional attention. Upon varying the redshift over the range considered later (the redshift varying between $z=0.13$ and $z=0.91)$, the amplitude of power spectrum varies mostly by amplitude not shape, with the former changing by $\mathcal{O}(2-3)$ \cite{Piras:2023aub}. }

\begin{figure}[ht]
    \centering
    \includegraphics[width=\linewidth]{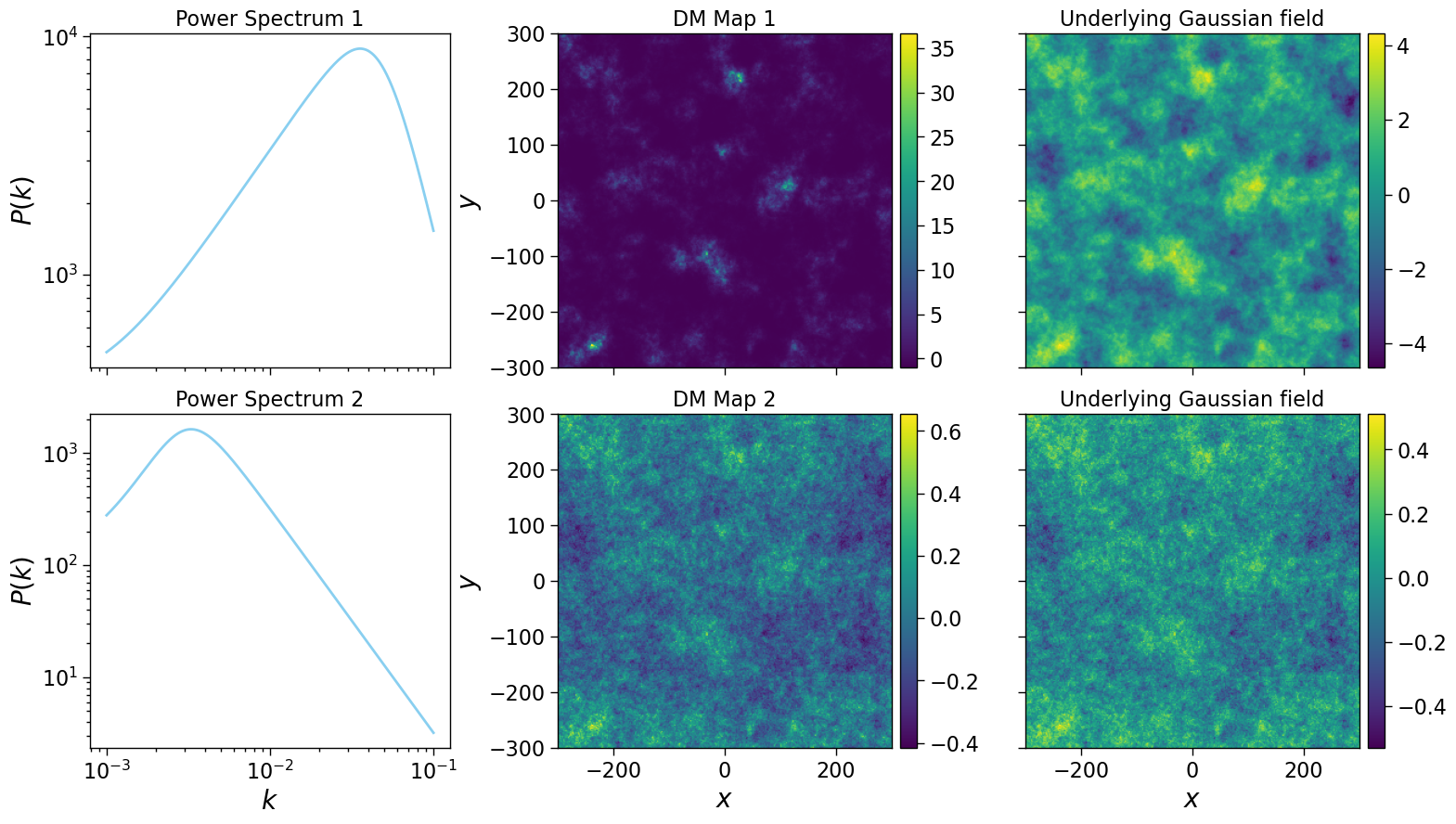}
    \caption{Comparison for two fiducial power spectra (\textit{left}), one realization of the respective log-normal field, modeling the DM density contrast (\textit{center}), and its underlying Gaussian random field (\textit{right}). 
    The latter two fields are related through Eq.~\eqref{eq:lognormal_transformation}. 
    The top power spectrum is peaked at large $k$ (small scales), resulting in small-scale structures, and vice versa for the second power spectrum, resulting in large-scale correlations. 
    Both power spectra were generated via Eq.~\eqref{eq: def phenomenological power spectrum}, with (1) \(\PA = 10^7\), \(\PnOne = 1\), \(\PnTwo = 4\), \(\PkEq = 2 \times 10^{-2}\), and \(\Pxi = 30\) and (2) \(\PA = 10^{12}\), \(\PnOne = 3\), \(\PnTwo = 5\), \(\PkEq = 2 \times 10^{-3}\), and \(\Pxi = 10\) for the power spectrum parameters.
    While the right-hand side is qualitatively different from full N-body simulations, the sampling from log-normal fields is computationally significantly less costly while ensuring large-scale correlations for the DM density.
    }
    \label{fig:example log normal field}
\end{figure}

For the DM power spectrum we rely on a phenomenological form, given by 
\begin{equation}
\label{eq: def phenomenological power spectrum}
    \PDM(\kvec) = \PA \, \frac{\keff^{\PnOne}}{\Pxi + \left( \frac{\keff}{\PkEq} \right)^{\PnTwo}} 
    \,,
\end{equation}
where $\kvec$ is the (three-dimensional) wave vector and $||.||$ is the Euclidean norm. 
We show two examples of the power spectrum and their corresponding log-normal fields in two dimensions in Fig.~\ref{fig:example log normal field}.
Two reasons made us choose a phenomenological model for the DM power spectrum: (\textit{i}) our previous work \cite{Leyde:2024tov} found little impact of using a power spectrum that depends on cosmological parameters and (\textit{ii}) it is faster to calculate.  

Since we want to evaluate the galaxy number counts in spherical coordinates, we need the DM density contrast in these new coordinates. 
The technical details of this interpolation scheme (cartesian to spherical coordinates) are described in Appendix~\ref{appendix:spherical_interpolation}.

\subsection{Galaxy bias}
\label{subsec: bias}

Numerical simulations of galaxy formation show that galaxy density does not scale linearly with the \gls{dm} density contrast. 
Instead, whether a galaxy is formed is a complex process of gravitational dynamics, baryonic feedback, and hydrodynamic processes \cite{Kuhlen:2012ft, Reddick:2012qy, Zheng:2015iia, Chaves-Montero:2015iga, Zavala:2019gpq, deMartino:2020gfi, Kokron:2021xgh}. 
In this work, we adopt a simplified model by assuming that the \gls{dm} density contrast within each voxel is directly related to the galaxy formation rate via a function, slightly modified from \cite{Neyrinck:2013ezr}, given by
\begin{equation}
\label{eq:def galaxy bias}
    \ratezhp \deffrom
    \biasamp 
    \exp\left[ - \left( \frac{1 + \biascut}{1 + \densityDM} \right)^{\biasepsilon} \right]
    \left(1 + \densityDM\right)^{\biasalpha} \,,
\end{equation}
where $\ratezhp$ is the galaxy density (i.e.~galaxy number per redshift and sky position) at a given redshift, $z$, at a given sky position, $\skypos$.
As a reminder, this quantity is the expression defined in Eq.~\eqref{eq: def ratezhpM}, marginalized over the absolute magnitude.
We will add in the absolute magnitude dependence in Sec.~\ref{subsec: magnitude distribution}. 
Finally, the symbols $\biasamp,\biasalpha,\biascut$ and $\biasepsilon$ denote the phenomenological bias parameters. 
The functional form above aims to capture the non-linear relation between DM density contrast and galaxy formation with two components: (\textit{i}) a power-law scaling and (\textit{ii}) an exponential suppression. 
This ensures that the galaxy rate scales as a power of the local dark matter overdensity: a positive value of $\biasalpha$ implies that galaxies preferentially form in overdense regions. 
The exponential cut-off suppresses galaxy formation in highly underdense regions. When the local dark matter density is much smaller than the cut-off parameter $\biascut$, the argument of the exponential becomes large, and the galaxy formation rate is exponentially suppressed, preventing galaxy formation in voids. The sharpness of this transition is determined by the parameter $\biasepsilon$.
The bias parameters will be inferred, although we fix $\biasamp=1$ because this parameter is strongly degenerate with the overall amplitude of the \gls{dm} matter power spectrum. 

Up to an overall normalization, the draw of the DM log-normal model and a choice of bias parameters allow us to formulate the associated PDF of the galaxy rate, $\ratezhp$, namely
\begin{equation}
    \label{eq:pdf galaxies}
    \p(z,\skypos|\hyperparams) = \frac{\ratezhp}{\int \ratezhp \,\dd z\,\dd \skypos} 
    \,,
\end{equation}
which describes the spatial distribution of galaxies for this choice of hyperparameters.

\begin{figure}[!ht]
    \centering
    \includegraphics[width=\linewidth]{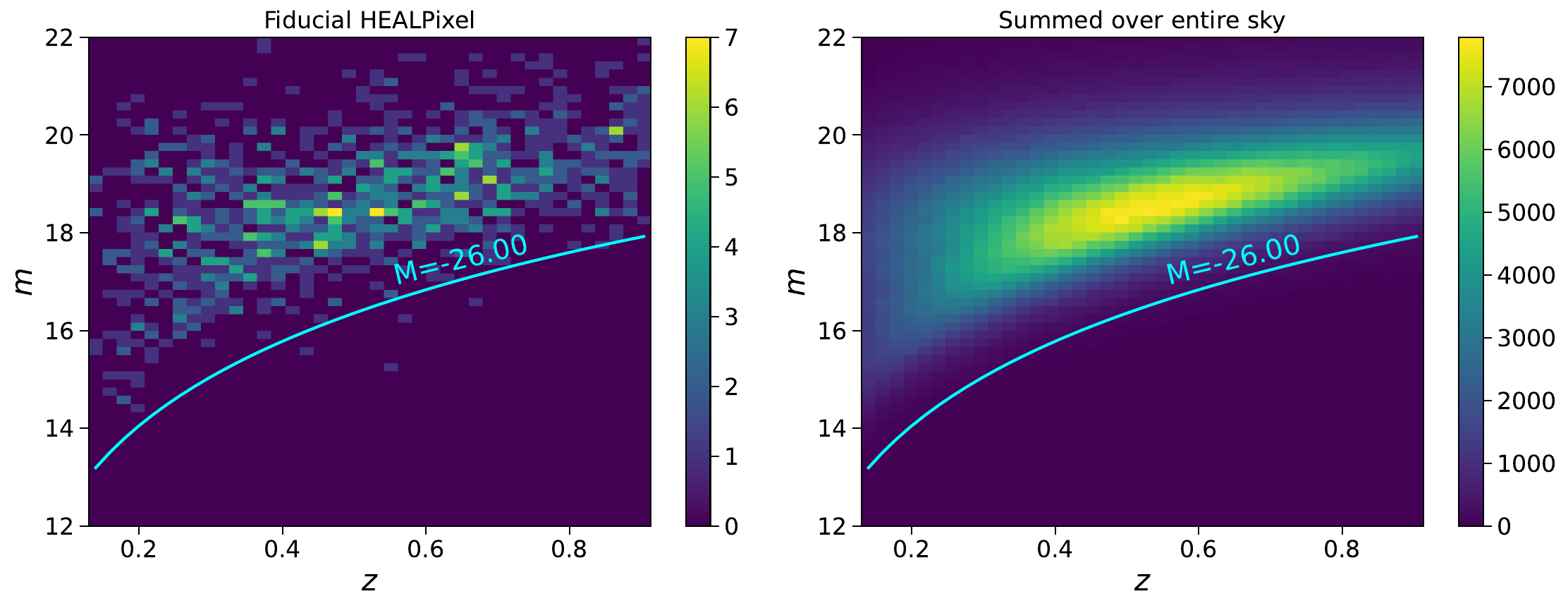}
    \caption{The apparent magnitude-redshift histogram for the (observed) simulated catalog, with (\textit{left}) a representative HEALPixel and the (\textit{right}) number counts summed over the entire sky. 
    To help guide the eye, we have added the apparent magnitude of a galaxy with fixed absolute magnitude (cf.~Eq.~\eqref{eq:def apparent magnitude}) in cyan, assuming the cosmology that was used to generate the catalog.
    Indeed, bright galaxies transition from low $m$ (at near-by redshifts) to larger $m$ for higher redshift, where the exact evolution depends on the cosmology through the luminosity distance-redshift relation.
    It is clearly apparent when the catalog becomes incomplete---at values of $m\gtrapprox 20$ the number counts plummet, a direct consequence of the detection probability that we simulate in Eq.~\eqref{eq:pdet modeling sigmoid}.
    We stress that the galaxy \textit{number count} is shown here (not galaxy density), resulting in a volumetric effect due to the increased volume of each pixel with increasing $z$. 
    While the representative HEALPixel (\textit{left}) is consistent with the overall structure of the distribution averaged over the sky (\textit{right}), the Poisson noise is larger. Indeed, the smoothness of the right figure is a result of recovering spatial homogeneity on large scales, not recovered at the individual HEALPixel level, as expected. }
    \label{fig:example mz plot millennium}
\end{figure}

\subsection{Magnitude distribution}
\label{subsec: magnitude distribution}

So far, we have only described the spatial clustering of galaxies, but we have not specified the distribution of the galaxy brightness.  
The brightness of a galaxy is quantified by its absolute magnitude, $M$. 
The apparent magnitude, $m$ is then a measure of the brightness of a galaxy as observed from Earth, related to the absolute magnitude and the luminosity distance, $d_L$, by the following equation
\begin{equation}
\label{eq:def apparent magnitude}
    m \deffrom 
    M + 5  \log_{10}(d_L) + 25 \,.
\end{equation}
In order to assess the completeness of the catalog, the distribution of galaxy absolute magnitudes is indispensable. 
If the ratio of bright to faint galaxies is known, at large distances, where solely the bright galaxies are detected, one can use those (in addition to this ratio) to ``calibrate'' the faint number of galaxies. 

At first glance, there seems to be a strong degeneracy between the measurement of the magnitude distribution and the estimation of the missing galaxies. However, as we will now argue, the high completeness of the survey at low redshift can break this degeneracy. 
%
Indeed, nearby voxels (low redshift) inform on the absolute magnitude distribution, up to a limit that corresponds to the faintest galaxy that can be observed near-by. 
As we move to larger redshifts (cf.~Fig.~\ref{fig:example mz plot millennium}), the apparent magnitude distribution shifts (following the \revvone{cyan} line of Fig.~\ref{fig:example mz plot millennium}) to larger values (i.e.~galaxies appear at higher apparent magnitudes for a fixed absolute magnitude).
For voxels at large redshift a significant portion of galaxies is too faint to be observed. However, from the inferred magnitude distribution at low redshifts and the abundance of bright galaxies we can infer the number of faint galaxies missed. 
From this argument, it is clear that the redshift (in)dependence of the magnitude distribution is an important systematic for the reconstruction. 

In order to be agnostic, we adopt a flexible magnitude distribution that does not rely on a specific functional form (as opposed to, for example, a Schechter distribution \cite{Schechter:1976iz}). 
This flexible approach is based on a Gaussian random field, which necessitates the assumption of a two-point function which determines how strongly and how fast the PDF fluctuates in magnitude space.
We detail the implementation of this approach in Appendix~\ref{app:magnitude_distribution} and show examples of possible magnitude distributions in Fig.~\ref{fig: gaussian_magnitude_distribution}. \revvone{Let us stress here that this distribution is assumed to be redshift-independent and independent from the DM density contrast.}

Our approach is to discretize the rate over both the spatial domain (redshift and sky position) as well as the magnitude space. 
As a consequence of this the corresponding reconstructed distributions cannot be trusted at scales below this discretization scale. 
We will describe the discretizations in the following Sections~\ref{subsec:redshift binning}, \ref{subsec:apparent magnitude binning} and \ref{subsec:healpix binning}: we count galaxies as they fall into a specific redshift, sky position and magnitude bin.

For the later reconstruction, we will also compute the galaxy Poisson rate marginalized over the apparent magnitude, i.e.~$\ratezhp \deffrom \int \ratezhpM \dd M$.
There are many objects at faint magnitudes, but because they are faint and are thus expected to have formed a small number of stars (i.e.~\gls{cbc} progenitors) their probability of hosting a GW event is believed to be small \cite{Giacobbo:2017qhh, Artale:2019tfl}. 
Therefore, we limit the reconstructed number counts to $M\leq \Mthreshold$, i.e.~$\ratezhp \deffrom \int_{-\infty}^{\Mthreshold} \ratezhpM \dd M$.
In the later results section we determine the threshold absolute magnitude value above which the reconstruction cannot be trusted. 
We anticipate here, that, unsurprisingly, under the assumption that the absolute magnitude distribution is independent of redshift, $\Mthreshold$ is determined by the faintest galaxies that are detected at low redshift.

The number counts inferred as described above depend on the cosmology, given that $\Mthreshold(z;\hyperparams)$ is cosmology-dependent. 
Despite this dependency, the reconstructed galaxy number counts exhibit only a weak dependence on the assumed values of $H_0$ over the considered prior, $\mightchange{\p(H_0)=\mathcal{U}(60\,\hu,80\,\hu)}$.

\subsection{Discretization scheme}

Throughout this work, we discretize the redshift, sky position and apparent magnitude, because it makes it more efficient to handle large datasets. 
The following section details the resolution of the discretization scheme in these coordinates.

\subsubsection{Redshift binning }
\label{subsec:redshift binning}

We histogram the galaxies into redshift bins of equal size according to their maximum-likelihood redshift. 
Typically, we start the reconstruction at a redshift of \mightchange{$0.1$} reach a maximum redshift of \mightchange{$0.9$}, with \mightchange{$\mathcal{O}(40)$} equal-size steps. 
In our numerical implementation, we generate the DM field on a cartesian grid. We then histogram the cartesian values in the spherical three-dimensional voxels and require each spherical voxel to contain at least one cartesian point. 
Therefore, we do not reconstruct the catalog at redshifts below the given minimum of \mightchange{0.1}, since this would require the cartesian grid of the DM density to be extremely fine (above current computational limits). 
The spherical discretizations chosen implies that the DM density is generated on a cube of \mightchange{$\mathcal{O}(100)$} segments per dimension.

\subsubsection{Apparent Magnitude Binning}
\label{subsec:apparent magnitude binning}


As mentioned previously, we reconstruct the number of galaxies in a given three-dimensional spatial bin and a given magnitude bin, allowing for a more informative assessment of the GW host distribution. 
In addition to spatial binning, we thus histogram galaxies based on their apparent magnitudes in $\mathcal{O}(60)$ equal-sized bins. 
Contrary to the redshift, we assume that the apparent magnitude is perfectly measured, an assumption that can be straightforwardly relaxed in the future.

\subsubsection{HEALPix binning}
\label{subsec:healpix binning}
We utilize the HEALPix (Hierarchical Equal Area isoLatitude Pixelization) framework \cite{Gorski:2004by, Zonca:2019vzt} to divide the celestial sphere into equal-area pixels. 
The HEALPix parameter \texttt{nside} controls the number of two-dimensional segments via $\# \text{pixels} = 12\, \texttt{nside}^2$, i.e.~the map resolution.
For our analyses, we choose \mightchange{$\texttt{nside}=16$}, which divides the sky in \mightchange{3072} segments, each with an area of $\sim 13\,{\rm deg}^2$.  
While it is desirable to have a higher resolution, we are limited by the cartesian grid used to generate the DM density contrast. For a simulation box of size $\mathcal{O}(6000)$~Mpc, we require $\mathcal{O}(10^2)$ grid points per dimension for \mightchange{$\texttt{nside}=16$}, which is $\sim 10^6$ pixels, i.e.~parameters to be estimated. 
In principle, the inference algorithm could handle larger parameter spaces but the model size is currently limited by the  GPU memory. Future multi-GPU inference development will thus allow to improve the sky discretization to higher $\texttt{nside}$.

Overall, we hence produce a three-dimensional (the HEALPix coordinates is one index for right ascension and declination) histogram of galaxies, amounting to a total of \mightchange{$\mathcal{O}(40(z)\times 3000(\skypos)\times 60 (m) \approx 7\times 10^6)$} bins.  
For an example of such a histogram for one given pixel in the sky, see the left panel of Fig.~\ref{fig:example mz plot millennium}. 

We can formulate the galaxy Poisson rate in apparent magnitude, which is related to the rate of Eq.~\eqref{eq: def ratezhpM} (and using Eq.~\eqref{eq:def apparent magnitude}) through
\begin{equation}
    \ratezhpm = \ratezhpMofzandm \,.
\end{equation}

\subsection{Redshift uncertainties}
\label{subsec: redshift uncertainties}

Measurement errors introduce uncertainties in the redshift, which are especially large for photometric redshifts.
This uncertainty is reflected in the number counts since we bin the galaxies in redshift by their (noisy) maximum-likelihood value. 
Here we will assume that the redshift uncertainty is Gaussian, with a standard deviation $\sigma_z$.
For $\hat z$ and $z$ the measured and true redshift, respectively, we have
\begin{equation}
\label{eq:gaussian redshift errors}
\p(\hat z | z) = 
    \mathcal{N}(\mu=z, \sigma = \sigma_z; x=\hat z)
\,,
\end{equation}
where $\mathcal{N}$ is the Gaussian distribution, with the standard convention for $\mu,\sigma$ and its argument $x$. 
Hence, the Poisson rate of the number counts at ($\hat z, \skypos, M$) is given by
\begin{equation}
    \label{eq:redshift uncertainty rate}
    \ratezhathpm = \int \ratezhpm \,\p(\hat z | z) \, \dd z \,,
\end{equation}
i.e.~a convolution of the Poisson rate with the Gaussian distribution.
This introduces features in the spatial distribution that point along the line of sight, and the length of these features is determined by $\sigma_z$.
This can be seen in the top right panel of Fig.~\ref{fig:slice phi} where we show a slice with constant azimuthal angle of the observed galaxy density binned in $\hat z$ and polar angle.

Note that we introduce a correction term that counter-acts the effects at the boundary of our redshift range for the numerical evaluation of Eq.~\eqref{eq:redshift uncertainty rate}. This correction enforces that a constant $\p(z)$ results in a constant $\p(\hat z)$. Otherwise, boundary regions would have lower probability density, due to the absence of galaxies outside the simulated redshift domain that would otherwise scatter into it, leading to an apparent deficit at the boundaries.

\subsection{Detection probability}
\label{subsec: detection probability}

The detection probability of galaxies is a key ingredient in the reconstruction of the full galaxy field. 
While this is a complicated function of the data (e.g.~the image, or spectrum), the classification as ``galaxy'' or ``not galaxy'' is typically approximated as a function dependent on the observed fluxes, possibly in multiple bands. 
We take the detection probability here as a function of the apparent magnitude in one given band and the sky position. 
Since these two quantities are not enough to determine with certainty whether the galaxy is detected or not, the detection probability becomes a continuous function between $0$ and $1$.\footnote{While the inclusion of the data point is a binary decision of the data, it is a continuous function as a function of its true parameters, since one marginalizes over different noise realizations. }

Formally, the detection probability for an observed galaxies with data $(\hat z, \skypos, m)$ is defined by 
\begin{equation}
  \p({\text{det}}|\hat z, \skypos, m)
  \deffrom
  \int 
  \p\left( \text{det} | \hat z, \skypos, m,\lambda  \right) \p(\lambda|\hat z, \skypos, m) 
  \dd \lambda
  \,,
\end{equation}
where ``det'' stands for ``detected'' and $\lambda$ is a set of galaxy parameters that is not included in our survey data. 
While the detection probability could depend on a wider set of galaxy parameters $\lambda$ we assume that the detection probability can be effectively modeled solely of the sky position $\skypos$, and apparent magnitude $m$, i.e.~$\p({\rm det}|\skypos, m)$.

In this work, we construct a flexible model for detection probability (detailed in App.~\ref{app: cumulative gaussian random field model for detection probability}). 
In order to understand the limitations when inferring the detection probability, let us recall that we have the schematic relationship $\p(\det) \sim \countGalaxiesObs / \countGalaxiesTrue$, with $\countGalaxiesObs$ and $\countGalaxiesTrue$ the observed and true galaxy counts, respectively. 
In order to infer $\p(\det)$, given the data $\countGalaxiesObs$, we thus have to know $\countGalaxiesTrue$. At first this seems to be circular, since we are interested in inferring the true number counts in the first place. 
Clearly, there is a strong degeneracy between the detection probability and the true number of galaxies---to account for a given number of observed galaxies one can simultaneously increase the true number of galaxies and decrease the detection probability, or vice versa. 
However, two assumptions can help break this degeneracy. 
At low redshifts, completeness is generally very high, thus providing a reliable estimate of the true galaxy number count.
Furthermore, the galaxy density averaged over large scales is roughly constant. 
Hence, if the detection probability is universal over large patches of the sky we can estimate it, schematically via $\p(\det) \sim \sum_I \countGalaxiesObspixel / \sum_I \countGalaxiesTruepixel$, where the index $I$ runs over all voxels in that patch with similar survey depth.
For sufficient homogeneous patches, we can thus jointly reconstruct the detection probability as a function of apparent magnitude along with the true galaxy Poisson rate. 

In practice, the detection probability as a function of apparent magnitude is not constant over the sky. 
While one could be tempted to reconstruct the detection probability independently for each sky pixel, we find this impossible due to the strong degeneracies described above.
For one sky pixel the true galaxy number count cannot be reliably estimated because of the cosmic web, i.e.~the Universe is not homogeneous on small scales. Indeed, the large-scale structure is why galaxy catalogs are useful for GW cosmology in the first place. 
(Remember that a redshift-sky position prior that is uniform in comoving volume leads to uninformative constraints on $H_0$).
However, we rely on the assumption that even though different sky pixels have different depths to which they can detect galaxies, the overall shape of the detection probability across the sky can be well-modeled by a universal function\footnote{This function will depend on the specific survey. }. Let us explain this in more detail. 

Say, one defines a characteristic apparent magnitude threshold, $m_{\rm thresh}$, for each pixel s.t.~$\p({\rm det}|m=m_{\rm thresh})=0.5$. This value will be high for pixels with deep coverage and low for shallow-coverage pixels. However, we can expect that the detection probability is similar for $m=m_{\rm thresh}(\skypos) + \Delta m$, independent of the pixel. 
We thus introduce a new variable $ X(m,\skypos) \deffrom m_{\rm thresh}(\skypos)-m$, where $m_{\rm thresh}$ is a characteristic apparent magnitude that depends on sky position, allowing us to connect sky pixels with different survey depth.
Throughout, we assume the function $X(m,\skypos)$ to be known. 
Then, rather than modeling the detection probability as a function of apparent magnitude, we model it as a function of $X$. This function $\p({\rm det}|X)$ will later turn out to be a sigmoid for the simulated data (see Fig.~\ref{fig:pdet estimate vanilla}), that we recover with a flexible distribution, i.e.~a Gaussian process.
For technical implementation details, consider App.~\ref{app: cumulative gaussian random field model for detection probability}. Of course, there is nothing fundamental about this choice and other distributions are possible, as long as they are generic enough to capture the shape of the true detection probability. 

For future data, inferring the detection probability might not even be necessary---one controls the detection algorithms and hence, the detection probability is known. 
One can simulate a large number of possible observations and run the detection algorithm on these synthetic datasets. The detection probability as a function of apparent magnitudes (possibly in different bands) and sky position can then be estimated to high precision.

\subsection{Likelihood}
\label{sec: likelihood}

Putting Eq.~\eqref{eq: def ratezhpM}, \eqref{eq:def galaxy bias} and \eqref{eq:redshift uncertainty rate} together, in combination with the modeling of the detection probability, we can combine the relevant terms of the above sections to compute the rate of detected galaxies \revvone{at a given} measured redshift $\hat z$, a sky position $\skypos$, and an apparent magnitude $m$ as
\begin{align}
 \nonumber
    \ratedet 
        =& 
    \underbrace{\rateAbs}_{\text{Overall galaxy rate}} 
        \times
    \underbrace{\p({\rm det}|f=f(m,\skypos))}_{\text{Detection prob.}} 
    \\
    &
        \times
        \int 
    \underbrace{\p(\hat z|z)}_{\text{Redshift uncertainties}} 
        \times
    \underbrace{\p(z,\skypos|\hyperparams)}_{\text{Spatial distribution}} 
        \times
    \underbrace{\p(M(m,z)|\hyperparams)}_{\text{Magnitude distribution}}
    \,\dd z
    \,.
\end{align}
This rate appears in the Poisson likelihood via
\begin{equation}
    \label{eq: poisson likelihood approx}
    \p(\nobsbinhat | \hyperparams) 
    =
    \frac{\exp
        \left[- \int \ratedet
            \, \dd z 
            \, \dd \skypos 
            \dd M
        \right] }{\prod_{\hat I \in \mathcal{I}}\nobsbinhat!}
     \prod_{\hat I \in \mathcal{I}} \ratedet ^\nobsbinhat
    \,,
\end{equation}
where the product runs over all bins, the set of which we denote as $\mathcal{I}$.

\subsection{Approximation of the number counts with the galaxy rate}
\label{subsec: approximation poisson rate}

In an ideal scenario, one would like to obtain a posterior in $n_{{\rm g},I}$, the true number galaxy number count in a bin $I$, i.e.~in a given redshift, sky position and absolute magnitude. 
Of course, the probability of finding a GW host at a given bin $I$ is proportional to the galaxy number realized, $\ngtruebin$, not the expected rate, $\ratezhpm$. 
However, obtaining posterior samples in discrete variables is computationally involved, and throughout we approximate the realized true number count with the Poisson rate $\ratezhpm$.
The standard deviation of the number count for a Poisson process is given by the square root of the mean, hence, we are making a relative error of $1 / \sqrt{\ratezhpm{}}$ in the number count.
Ultimately, for GW cosmology, we are interested in $\p(z,\skypos)$ because redshift information conditioned on sky position is needed. 
Since the number counts $n_{{\rm g},I}$ summed over all absolute magnitude bins are of the order of $10^2$, we can estimate the relative error due to this approximation to be $\mathcal{O}(1/\sqrt{100}=0.1)$.

\subsection{Sampling}
\label{subsec:sampling}

Generating posterior samples from a high-dimensional posterior distribution is a challenging task because the posterior tends to occupy a small volume of the overall prior volume. 
For this reason we make use of Hamiltonian Monte Carlo (HMC) sampling, which proposes new samples of an MCMC chain by integrating trajectories through the space of parameters, ensuring proposed samples to have high acceptance probability. The integration of paths requires taking gradients, which we can easily compute, as our model is implemented using the auto-differentiable programming package \jax{} \cite{jax2018github}.

We generate our samples using a particular type of HMC, the No-U-turn sampler (NUTS) \cite{Neal:2011mrf, Hoffman:2011ukg, Betancourt:2017ebh} that is implemented in the \numpyro{} package \cite{Phan:2019elc}. NUTS extends HMC to generate fairly uncorrelated samples by identifying a U-Turn in the trajectories of the HMC sampling process. No burn-in removal is required for this sampler as this accounted for by the so-called warm-up phase. 
During this phase (for which we choose a length of 2000 samples), one tunes the step-size and mass matrices of the integrator. We then draw 1500 posterior samples of our target parameters with \mightchange{three} chains.

\section{Simulated data}

\label{sec: simulated data}

\begin{figure}[ht]
    \centering
    \includegraphics[width=1\linewidth]{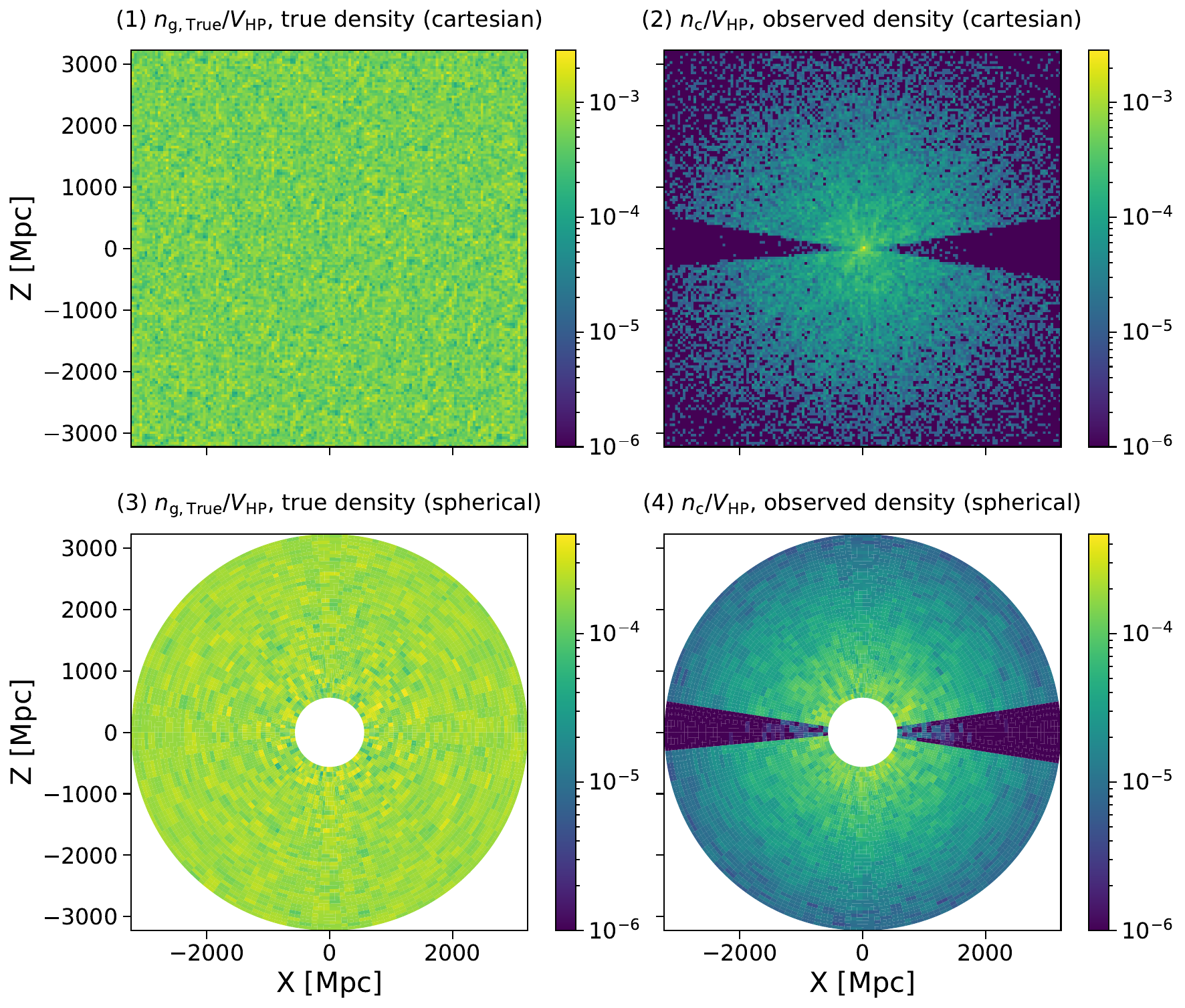}
    \caption{A slice through the stacked Millennium simulation illustrating the three-dimensional (spherical) pixelization. (1) The true galaxy density in cartesian coordinates. (2) The observed galaxy density in cartesian bins. (3) The true galaxy density in spherical bins. (4) The observed galaxy density in spherical bins. Although the spherical discretization happens in redshift space, we have assumed here a fiducial cosmology of \mightchange{$H_0=67\,\hu$} to plot the volumes in comoving volume (panels 3 and 4), for illustration only.
    Also note that we mimic the observational effect of the milky-way by imposing a significantly lower apparent magnitude threshold in $\mightchange{400}$ out of the \mightchange{3072} sky pixels, in the plane of the milky-way disk.
    This leads to the cone structure in which virtually no galaxies are observed. 
    Panel 3 shows larger densities near the center of the plot. This is due to the three-dimensional voxel volumes being smaller which in turn means larger variance in the galaxy number counts and hence, estimated galaxy density. 
    The volumes that are located at further distance from the center average over larger volumes, have lower variance and thus appear smoother. 
    } 
    \label{fig:slice phi}
\end{figure}

To validate our method on realistic data, we use the mock galaxy catalogs of \cite{Bertone:2007sj}, which are based on the Millennium Simulation \gls{dm} density field \cite{Springel:2005nw}.
The original catalog represents a cubic volume of $ (500 \Mpc/h)^3 $. 
Since we want to build a comparable catalog to \gladeplus{} \cite{Dalya:2021ewn}, this coverage is insufficient for our analysis (since \gladeplus{} covers a much larger volume), so we replicate the catalog in an $ 9 \times 9 \times 9 $ tiling (including random rotations), effectively increasing the total volume to $ (4500 \Mpc/h)^3 $, while maintaining the large-scale statistical properties of the distribution.
For this tiling, we use the same snapshot of the Millennium simulation at $z=0$.
To match the overall number of observed galaxies in \gladeplus{}, we select only $1\%$ of all available galaxies.
This yields approximately a total of $\mathcal{O}(3\times 10^6)$ galaxies. 
Assuming a $\Lambda$CDM cosmology with $\mightchange{H_0=67\,\hu}$ and $\mightchange{\Omegam=0.3}$, we convert the galaxy comoving coordinates to redshift and sky position. 
\revvone{The galaxy absolute magnitudes are taken directly from the simulation of \cite{Bertone:2007sj}. }

We apply Gaussian uncertainties on the redshift measurement, cf.~\eqref{eq:gaussian redshift errors}, with the standard deviation 
\begin{equation}
\label{eq: redshift error simulated}
    \sigma_z(z)
    =
    0.01 \times (1 + z) \,,
\end{equation}
weakly dependent of the redshift, and independent of sky position.
The galaxies are then histogrammed according to their measured redshift value $\hat z$, which introduces features along the line-of-sight, faintly noticeable in Fig.~\ref{fig:slice phi}, \mightchange{panel 2}. 
While our proposed method can account for redshift uncertainty as a function of the sky position, it requires a priori knowledge of the uncertainty of the redshift measurement.
Sec.~\ref{sec: results} will test various mis-modeling effects including under- and over-estimating the redshift uncertainties. 
Throughout, it is assumed that the sky position is perfectly known. 

We impose a probabilistic selection function that depends on a characteristic magnitude, $\mudet$ and a characteristic fall-off scale, $\sigmadet$, through 
\begin{equation}
\label{eq:pdet modeling sigmoid}
    \p({\rm det}|m) = {\rm sig}\left(\frac{\mudet - m}{\sigmadet}\right) \,,
\end{equation}
where the sigmoid function was defined in Eq.~\eqref{eq: def sigmoid}. The above functional form of the detection probability is only assumed for data generation, for the reconstruction we use the Gaussian process model of App.~\ref{app: cumulative gaussian random field model for detection probability}. 
Although detection algorithms are typically deterministic (not probabilistic as the above equation implies), this description attempts to capture the impact of other variables (such as the flux in other bands) on the galaxy detection probability. 

In the present work, we choose \mightchange{$\mudet=19$ and $\sigmadet=0.6$}, resulting in the fall-off of number counts appearing in Fig.~\ref{fig:example mz plot millennium}---very few galaxies above an apparent magnitude of $\mudet + 3\times \sigmadet \approx 20.8$ are detected. To model the milky-way disk, we impose a smaller apparent magnitude of $\mudet=12$ for 400 out of the 3072 HEALPixels. We find that sampling is facilitated if we completely mask out these low-coverage HEALPixels.

\section{Results}
\label{sec: results}

Using the likelihood formulated in Eq.~\ref{eq: poisson likelihood approx} and following the sampling procedure described in Sec.~\ref{subsec:sampling}, we obtain posterior samples in the galaxy Poisson rate. 
The reconstructed galaxy Poisson rate is implemented in the code as a three-dimensional rate (redshift, HEALPix index, absolute magnitude), but represents a four-dimensional quantity (redshift, polar angle, azimuthal angle, absolute magnitude).
In order to visualize the results, we will either sum over the galaxy rate along a given variable (e.g.~integrating over the absolute magnitude) or keep the variable constant (e.g.~fixing the sky position), as specified in the figure captions.

In the following, we summarize the reconstruction results and test for different systematics.  
For this purpose, we will construct different statistics to evaluate the accuracy of the reconstruction with respect to the ``real'' Universe, i.e.~\mightchange{the Millennium simulation}. 
Throughout, we assume a redshift range from 0.13 to 0.91 in 40 equal-size steps. For the HEALPix discretization we are using $\texttt{nside}=16$, leading to 3072 sky pixels.
The spatial discretization schema implies the DM density contrast to be computed on a \mightchange{$112^3$} cartesian box.
The apparent magnitude is discretized from \mightchange{12} to \mightchange{22} in \mightchange{60} equal-size steps.
Finally, the detection probability is computed over a grid of 200 bins in the range of $\Xnorm\in[-5,5]$ (cf.~Eq.~\eqref{eq:def Xnorm}).

\subsection{Vanilla case}
\label{sec:result vanilla}

\begin{figure}[p]
    \centering
    \includegraphics[width=\linewidth]{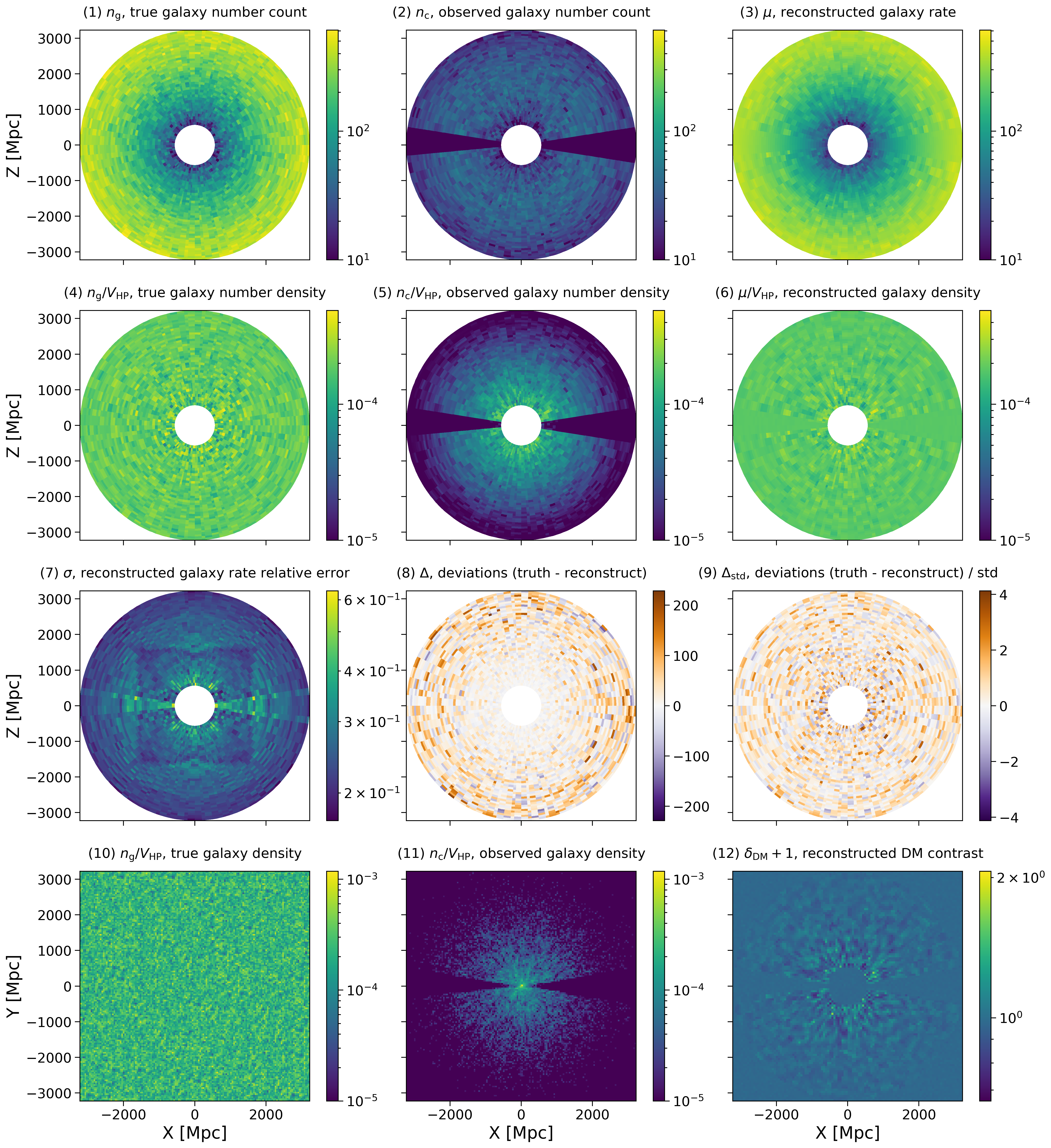}
    \caption{Slice through constant $\phi$ (azimuthal sky angle). The numbered panels are described in the main text. We denote the pixel volume by $V_{\rm HP}$.
    To convert the rate in redshift space into comoving space we assume a reference cosmology with $H_0=67\,\hu$ and $\Omegam=0.3$. 
    }
    \label{fig: reconstruction phi-slice}
\end{figure}
For the reference result, we assume a prior of \mightchange{$\p(H_0 / (\hu))=\mathcal{U}(60,80)$} and set the Gaussian redshift uncertainty as $\sigma_z = \mightchange{0.02}$, independent of redshift.
The priors assumed for the remaining parameters are collected in App.~\ref{app: prior range}.
We jointly estimate the absolute magnitude distribution, the galaxy bias, the DM power spectrum, the DM density contrast, the selection bias and the overall rate of galaxies. 
We will contrast this result in later sections against fixing $H_0$ (Sec.~\ref{subsec: impact fixing H0}), changing the assumed redshift uncertainties (Sec.~\ref{subsec: impact redshift uncertainties}), and varying the absolute magnitude threshold for the faintest reconstructed galaxies (Sec.~\ref{subsec:impact Mthreshold}). 

\subsubsection{Validating the reconstruction (redshift + polar angle)}

We start with marginalizing over the absolute magnitude coordinate\footnote{We reconstruct the number counts up to an absolute magnitude of $\Mthreshold=\mightchange{-19}$. We determine in Sec.~\ref{subsec:impact Mthreshold} the absolute magnitude above which the reconstruction can no longer be trusted. }, and fix the azimuthal angle to 0 and $\pi$ in order to obtain the entire $x-z$ plane, shown in Fig.~\ref{fig: reconstruction phi-slice}. 
The panels are, from left to right, top to bottom: (1) The true galaxy number count, galaxies binned by their true redshift and polar angle, (2) the observed galaxy number count, binned by their measured redshift and polar angle, (3) the reconstructed galaxy Poisson rate in true redshift and polar angle, (4-6) the galaxy densities (same as panel (1-3) but dividing out the comoving volume), (7) the relative error of the galaxy Poisson rate, i.e.
\begin{equation}
\label{eq: def relative deviations}
   \text{Panel (7):}\quad \sigma \deffrom
   \frac{{\rm std}(\ratezhp)}{{\rm median}(\ratezhp)} \,,
\end{equation}
where ``std'' denotes the standard deviation.
Panel~(8): The deviation between the true galaxy number counts and the galaxy Poisson rate, 
\begin{equation}
\label{eq:def Delta deviation}
   \text{Panel (8):}\quad
    \Delta\deffrom\ngtruebin - \ratezhp
\,. 
\end{equation}
Panel~(9): To assess the deviations in units of the standard deviation, we define the relative deviations as
\begin{equation}
\label{eq:def Delta std}
   \text{Panel (9):}\quad
    \Deltastd \deffrom \frac{\Delta }{{\rm std}(\ratezhp)} 
\,. 
\end{equation}
Panel~(10): The true galaxy distribution in cartesian coordinates, panel~(11): the observed galaxy distribution in cartesian coordinates\footnote{In order to illustrate the result of the uncertain redshift we introduce a rescaled comoving coordinate for each galaxy, depending on the measured redshift for that galaxy. This is only for visualizing purposes and does not enter the analysis. }, panel~(12): the reconstructed DM density contrast in cartesian coordinates, $\densityDM + 1$.

The reconstructed galaxy Poisson rate (3) closely follows panel (1), we thus conclude that the reconstruction works well. 
To put this in quantitative terms, we have computed panel (8), that measures the difference between true and reconstructed galaxy number counts, cf.~Eq.~\eqref{eq:def Delta deviation}. 
We see that the difference fluctuates around zero, with the deviations increasing with redshift. For larger redshifts, the bins have larger comoving volumes and hence include more galaxies. 
To check the deviation in terms of the inferred uncertainty, we rescale the differences to obtain $\Deltastd$ (panel 9), defined in Eq.~\eqref{eq:def Delta std}. 
The deviations $\Deltastd$ are distributed homogeneously across all angles and redshifts, indicating no systematic bias in the reconstruction. 
We note a slight dominance of red pixels over blue ones, indicating that we tend to underestimate the true galaxy number count. 
This can be partially traced back to an overestimation of the detection probability, discussed below. 
In the later Fig.~\ref{fig: relative deviations histogram H0 impact} \mightchange{(yellow line)}, we histogram the relative deviations showing close agreement with the expected zero-mean unit-variance normal distribution.\footnote{In case of a Gaussian posterior these deviations should follow a zero-mean unit-variance Gaussian distribution. In our case there are two complications to this: (\textit{i}) the posterior is not Gaussian and (\textit{ii}) we compare the reconstructed Poisson rate against the true number counts (rather than the reconstructed true galaxy number count vs.~the true number counts). Both points will - even for an accurate model - shift the distribution of $\Deltastd$ from a zero-mean unit-variance normal distribution. }

Panel \mightchange{(7)} of Fig.~\ref{fig: reconstruction phi-slice} shows the relative deviations (cf.~Eq.~\eqref{eq: def relative deviations}) to decrease over redshift. 
This is a consequence of the relative error of a Poisson-distributed quantity to decrease with the number counts, and the number counts increasing over redshift due to the voxels' fixed angular size. 
A similar effect leads to increased relative error bars in the ``milky-way'' band: since we have masked out the low observed number count in this sky region we have little constraining information on the galaxy distribution here. 
Panel \mightchange{(12)} shows the DM density contrast is much better constrained for near-by regions, but reverts to the prior (i.e.~uniform) for distant voxels.
We also note that the finer discretization scheme of the smaller cube in the center is visible in \mightchange{(7)}, pointing to the necessity of a more sophisticated discretization scheme for future work. 

\begin{figure}[ht]
    \centering
    \includegraphics[width=\linewidth]{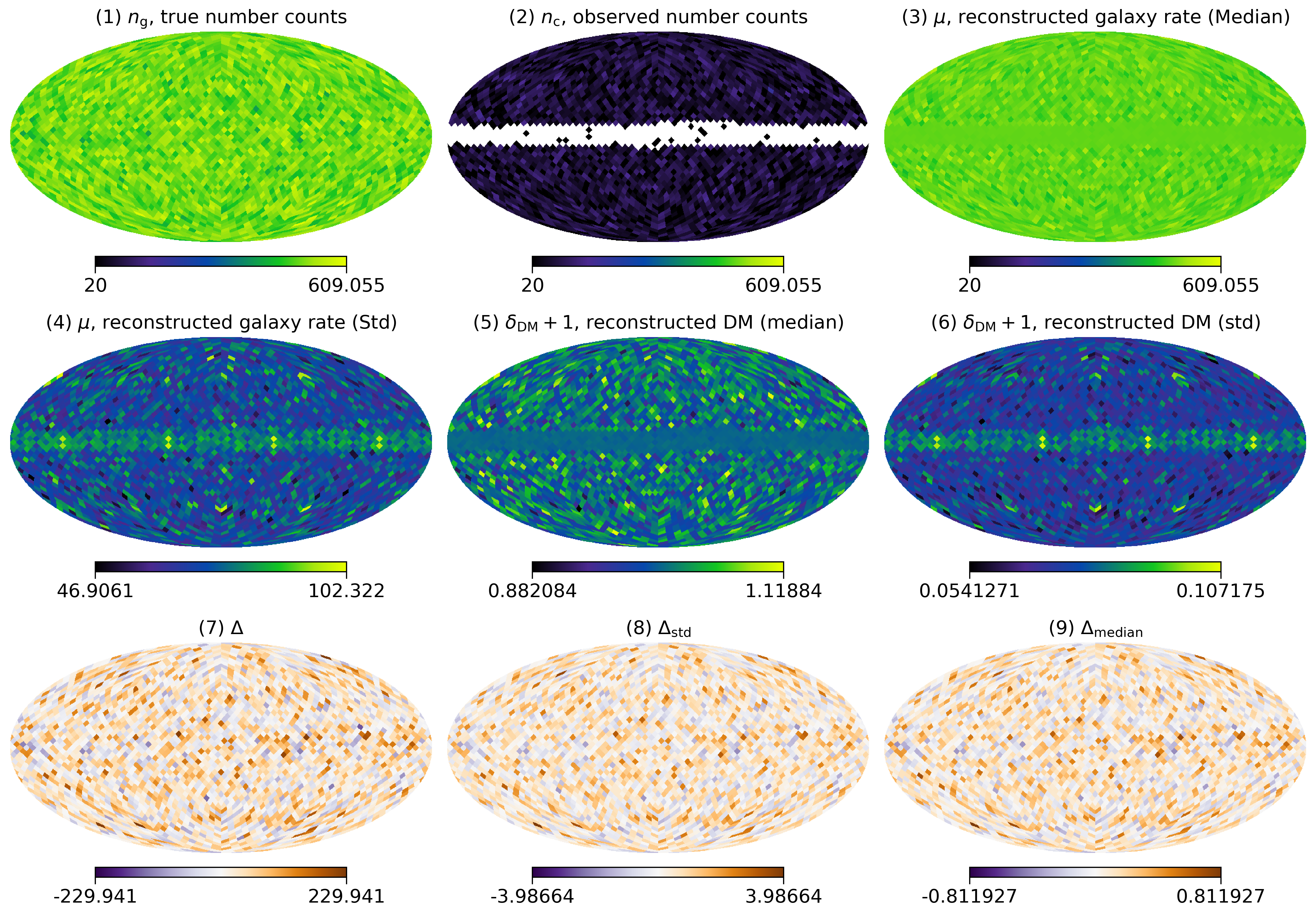}
    \caption{
        Collection of various quantities for a given redshift slice of $z\approx\zsliceplot$. 
        Panels (7)-(8) were defined in Equations~\eqref{eq:def Delta deviation}, \eqref{eq:def Delta std} and describe the deviations between the reconstructed galaxy Poisson rate and the true number counts; in absolute units (panel 7) and in units of standard deviations (panel 8). 
        Finally, Eq.~\eqref{eq:def Delta median} defines a new quantity called (relative) deviation, shown in panel (9).
        In order to simulate a milky-way like observation bias, we apply an apparent magnitude cut of \mightchange{12} in a horizontal band for \mightchange{400} HEALPixels. 
        }
    \label{fig:mollweide slice in redshift}
\end{figure}

\subsubsection{Validating the reconstruction (polar and azimuthal angle)}

The next representation of the results, Fig.~\ref{fig:mollweide slice in redshift}, marginalizes again over the absolute magnitude, but fixes the redshift to $z\sim\zsliceplot$. With this figure, we would like to illustrate the reconstruction for far-away regions and their sky position-dependence. 
Most of the panels are similar to the \mightchange{previous} Fig.~\ref{fig: reconstruction phi-slice}, with the exception of the last panel, panel (9), which shows the deviations in units of the Poisson rate, i.e.
\begin{equation}
    \label{eq:def Delta median}
    \Deltamedian \deffrom \frac{\Delta}{{\rm median}(\ratezhp)} \,.
\end{equation}
We note that the standard deviation of the estimated galaxy Poisson rate is notably larger in the band along the horizon where we have little data. 
This is expected, since a smaller detection probability leads to a larger uncertainty in the inferred Poisson rate.
All three types of deviations, $\Delta, \Deltastd$, and $\Deltamedian$, are distributed homogeneously across all sky pixels, indicating that no systematic bias in our reconstruction. 
The deviations in units of standard deviations reach up to $\sim 4.0$, compatible with (but slightly larger than) the expected value of deviations for Gaussian fluctuations (drawing 3072 values (i.e.~sky pixels) from a zero-mean, unit-variance normal distribution). 
The uncertainties of the inferred number counts are thus underestimated -- this is partially due to approximating the reconstructed number counts by the galaxy Poisson rate (cf.~Sec.~\ref{subsec: approximation poisson rate}). While the overall rate matches the true number counts, the reconstructed galaxy Poisson rate is more homogeneous than the true number counts ((3) vs (1)). 
Again, this is due to the aforementioned approximation. Additionally, the DM density description by a log-normal field, or the simple local bias prescription might not able to fully capture the complexity of the galaxy distribution.

\subsubsection{The inferred detection probability}

\begin{figure}[ht]
    \centering
    \includegraphics[width=0.65\linewidth]{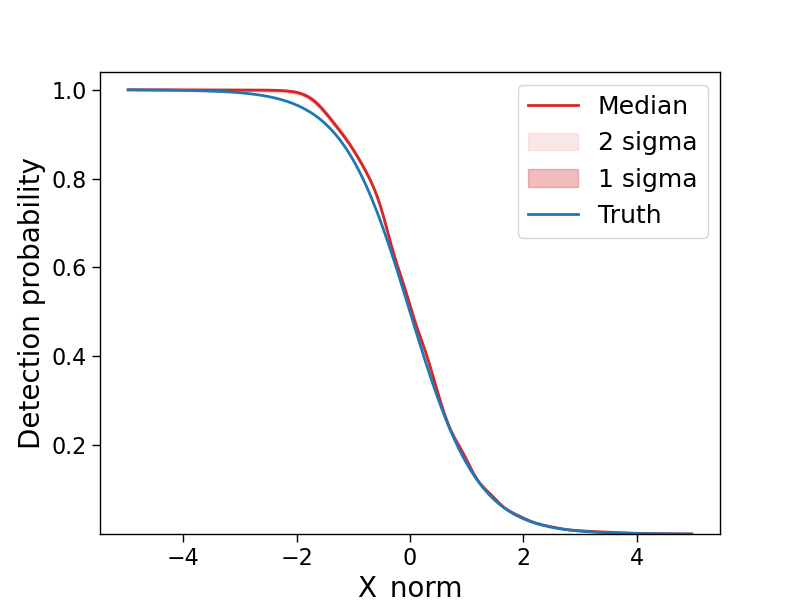}
    \caption{The estimated detection probability as a function of apparent magnitude. The normalized variable $\Xnorm$ (defined in Eq.~\eqref{eq:def Xnorm}, with $\mmean=0$ and $\mscale=1$) and is directly related to the apparent magnitude. 
    We reconstruct (blue) the true detection probability (red) with good precision, although the detection probability is overestimated for small (bright) magnitudes.
    }
    \label{fig:pdet estimate vanilla}
\end{figure}
In Fig.~\ref{fig:pdet estimate vanilla} we show our estimate of the detection probability as a function of $\Xnorm$ defined in Eq.~\eqref{eq:def Xnorm}. 
As we have elaborated above, because of unseen latent variables, this probability density takes continuous values between 0 and 1. 
The overall shape is well reconstructed, which is reflected in the faithful reconstruction of the catalog. 
However, the detection probability is slightly overestimated for small $\Xnorm$ (bright objects), for reasons not completely understood. One possibility could be the mis-modeling of the redshift uncertainties, where we approximate the redshift-dependent error (cf.~Eq.~\eqref{eq: redshift error simulated}) with a constant $\sigma_z$. 
We have also attempted to fit the detection probability as a universal function that is shifted in each sky pixel, the shift being estimated for each sky pixel independently. This leads to difficulties in sampling that originate from strong degeneracies between the detection probability and the true galaxy rate. 
(For a given observed galaxy number count one can augment the true galaxy number count while decreasing the detection probability, keeping the likelihood constant.)
For a real survey, we expect the detection probability to be known (since the observer uses fully-controlled detection algorithms). Hence, while a good sanity check here, robustly estimating the detection probability is not necessary for building completed catalogs.

\subsubsection{The reconstructed line-of-sight redshift prior}

\begin{figure}[htb!]
    \centering
    \begin{subfigure}[b]{0.7\textwidth}
        \includegraphics[width=\textwidth]{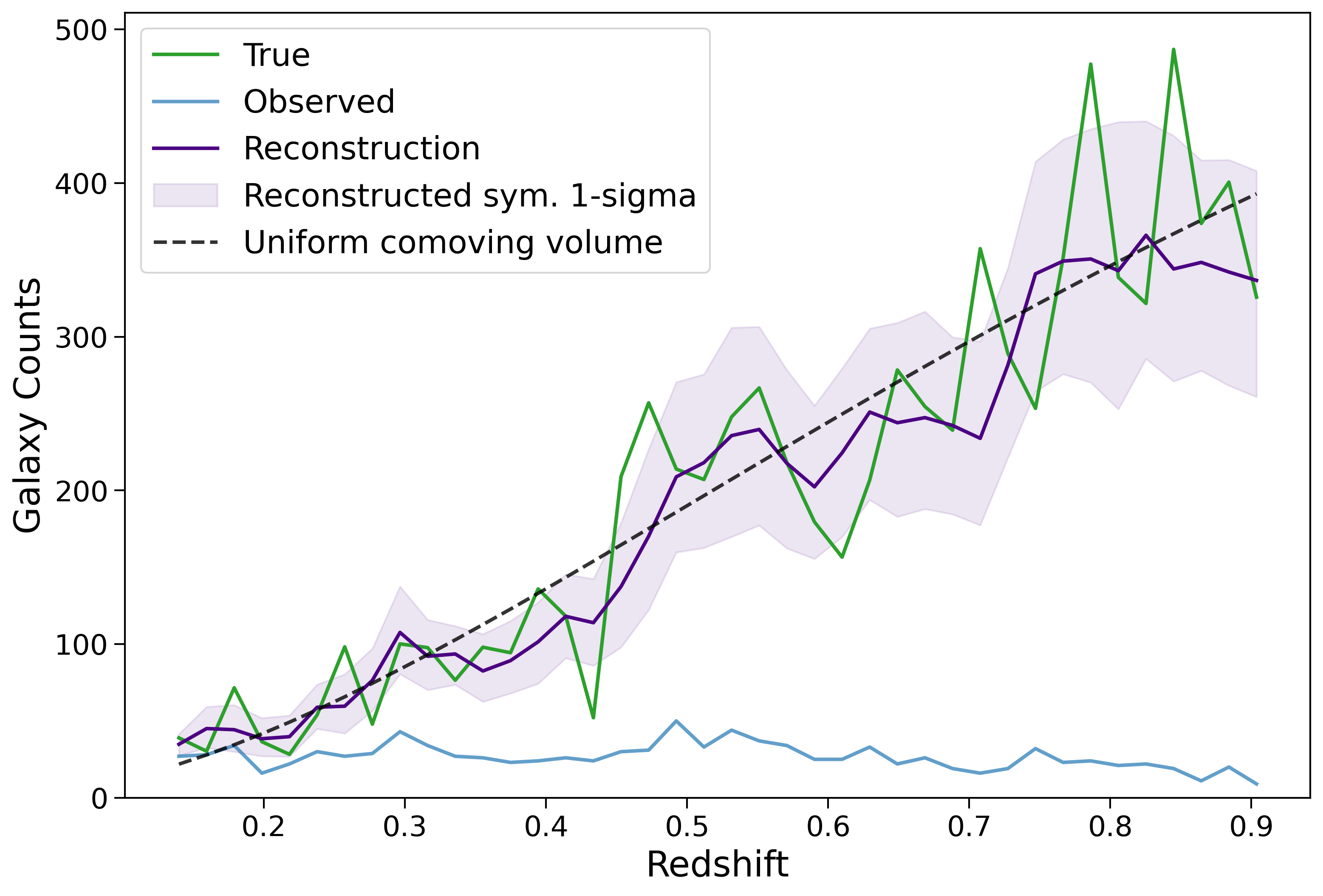}
        \caption{Fiducial HEALPixel with high completeness. }
        \label{fig:pixel122}
    \end{subfigure}
    \hspace{1cm}
    \begin{subfigure}[b]{0.7\textwidth}
        \includegraphics[width=\textwidth]{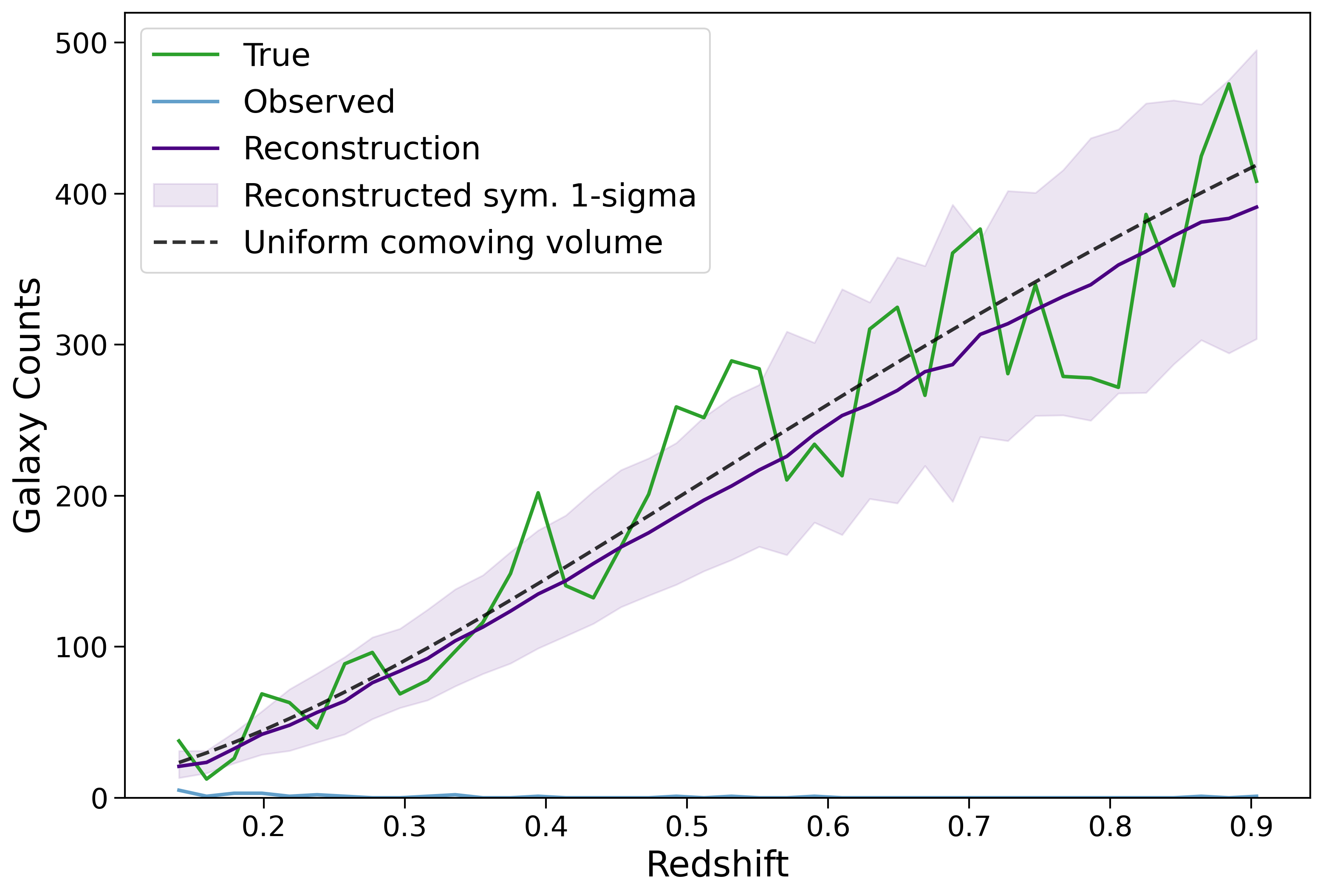}
        \caption{Fiducial HEALPixel with low completeness.}
        \label{fig:pixel1400}
    \end{subfigure}
    \caption{Galaxy number count for two representative HEALPixels. One has a deep apparent magnitude cut (\textit{top}), the other a shallow cut (\textit{bottom}).
    In the case of the ``deep coverage'' HEALPixel the reconstruction recovers the overall large-scale structure, whereas the empty HEALPixel recovers a galaxy rate following a distribution that is uniform in comoving volume. 
    Also note that the uncertainties increase as one approaches larger redshifts.
    For reference, we plot a uniform-in-comoving-volume distribution for the true cosmology (dashed line), normalized to the true galaxy number counts (\mightchange{green} line). 
    Recall that this uniform distribution yields an uninformative $H_0$ posterior, cf.~Sec.~\ref{subsec: motivation} and Fig.~\ref{fig:completeness impact schematic}. 
    }
    \label{fig:galaxy_counts_two_pixels}
\end{figure}
To understand the implications of the reconstruction, we plot the true galaxy number count and inferred galaxy Poisson rate for a given HEALPixel, summing over the absolute magnitude counts. 
We focus on two different cases: (\textit{i}) a pixel with a large number of observed galaxies, i.e.~with a large apparent magnitude threshold, and (\textit{ii}) a pixel in the ``milky-way region'', i.e.~with a small apparent magnitude threshold.
The two cases are shown in the top and bottom panels of Fig.~\ref{fig:galaxy_counts_two_pixels}, respectively.
There are two differences between the true and observed number counts: while they agree for low redshifts, the ratio of observed to true galaxy number counts drops for larger redshift due to fainter galaxies being missed.
The second difference is that the observed number counts lack the small-scale fluctuations of the true number counts---this is due to the redshift uncertainties smearing out any fluctuations smaller than the standard deviation of the redshift distribution.

In case of the more complete sky pixel, the inferred galaxy Poisson rate is in good agreement with the true number counts, following the fluctuations that are informed by the observed galaxies. 
We note however that outliers in the true galaxy number to high values are generally outside the 1-sigma intervals. 
This is partly due to our approximation of the true number counts with the galaxy Poisson rate which is generally smoother and has hence difficulties accommodating outliers. 
In the case of the less complete sky pixel, the inferred galaxy Poisson rate is much more homogeneous, closely following a uniform-in-comoving-volume distribution.
This is a direct consequence of the poor constraining power of the data about the galaxy distribution in this HEALPixel.

\begin{landscape}
\begin{figure}[p]
    \centering
    \includegraphics[width=24cm]{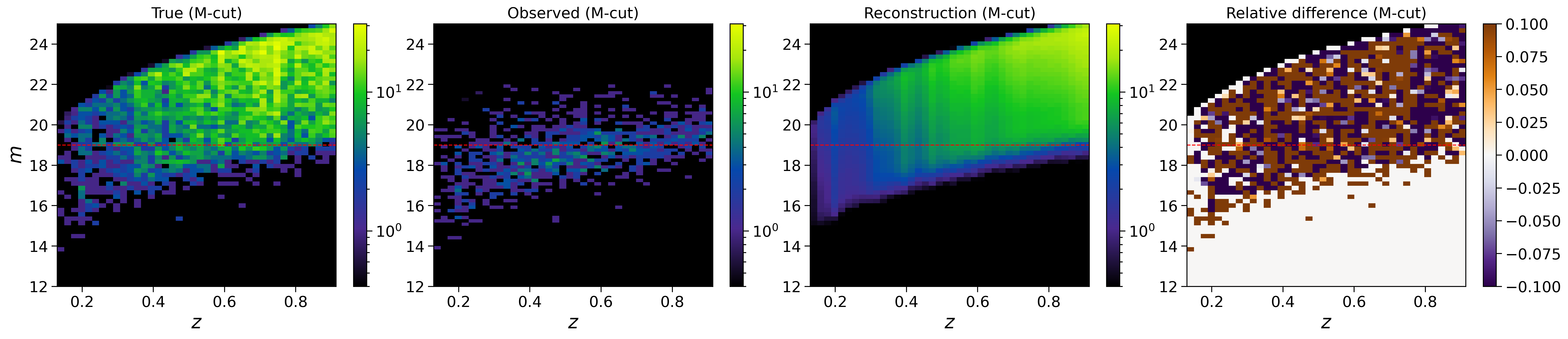}
    \caption*{Fiducial HEALPixel.}
    \vspace{1cm}
    \includegraphics[width=24cm]{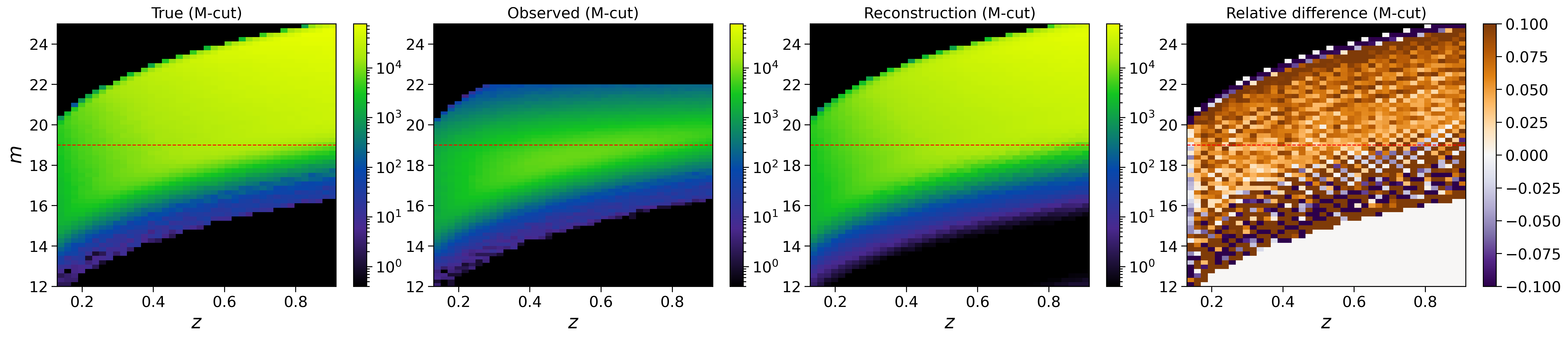}
    \caption*{Full sky (summed over all HEALPixels). }
    \caption{Galaxy number counts in apparent magnitude and redshift for a selected pixel (\textit{top}) and summed over the full sky (\textit{bottom}).
    The columns from left to right are: the true number counts, the observed number count, the reconstructed Poisson rate, the relative difference, i.e.~Eq.~\eqref{eq:def Delta median}. 
    The cut in the observed number counts at $m=22$ is due to the likelihood being evaluated on a grid up to this value -- very few galaxies are observed above these apparent magnitudes. 
    Note that the upper-left dark triangle is due to us excluding all magnitudes above \mightchange{$-19$}, while the lower-right section is dark because no galaxies above a certain brightness exist. During inference, we do not impose any upper threshold on the absolute magnitude, this is only applied in post-processing. 
    }
    \label{fig:galaxy_counts_zm}
\end{figure}
\end{landscape}

\subsubsection{The inferred absolute magnitude distribution}
So far, we have only shown the results as summed over the absolute magnitude dimension.
However, the reconstruction is strongly dependent on the absolute magnitude distribution, turning it in an important feature that has to be inferred from the data and that we now want to highlight. 
In Fig.~\ref{fig:galaxy_counts_zm} we show the reconstructed galaxy Poisson rate and various other quantities both for (\textit{i}) a given HEALPixel \mightchange{(top)} and (\textit{ii}) summed over the sky \mightchange{(bottom)}, in apparent magnitude and redshift bins. 
The first column shows the true galaxy number counts, the second column shows the observed galaxy number counts, and the third column shows the inferred galaxy Poisson rate.
Finally, the last column shows the relative deviation between the inferred galaxy Poisson rate and the true galaxy number counts, defined in Eq.~\eqref{eq:def Delta median}.

Let us first focus on the true catalog (first column).
We see the apparent magnitude evolution (the approximately square root trend), reminiscent of Eq.~\eqref{eq:def apparent magnitude} and Fig.~\ref{fig:example mz plot millennium}.
As galaxies are located at larger redshifts, they appear fainter (i.e.~at larger $m$). 
We also notice an overall increase in the number counts with redshift, which is due to the larger associated comoving volume for voxels at larger redshifts.
The second column shows the observed galaxy number counts, which are smaller than the true number counts due to the lower detection probability at larger apparent magnitudes.
We have marked the apparent magnitude at which only $50\%$ of the galaxies are detected as a red dashed horizontal line. Above this line fewer galaxies are detected -- and at an apparent magnitude of $m=22$ the observed catalog is essentially empty. 
Note that we have applied an absolute magnitude cut of $\Mthresh = \mightchange{-19}$, which is reflected in a dark region in the top left region of each plot. 
We decide to eliminate these galaxies (in post-processing) since they have low GW host probability and hence also contribute little to the redshift measurement for GWs.
We will explore the impact of this cut in more detail in Sec.~\ref{subsec:impact Mthreshold}.

\begin{figure}[ht]
    \centering
    \includegraphics[width=0.8\linewidth]{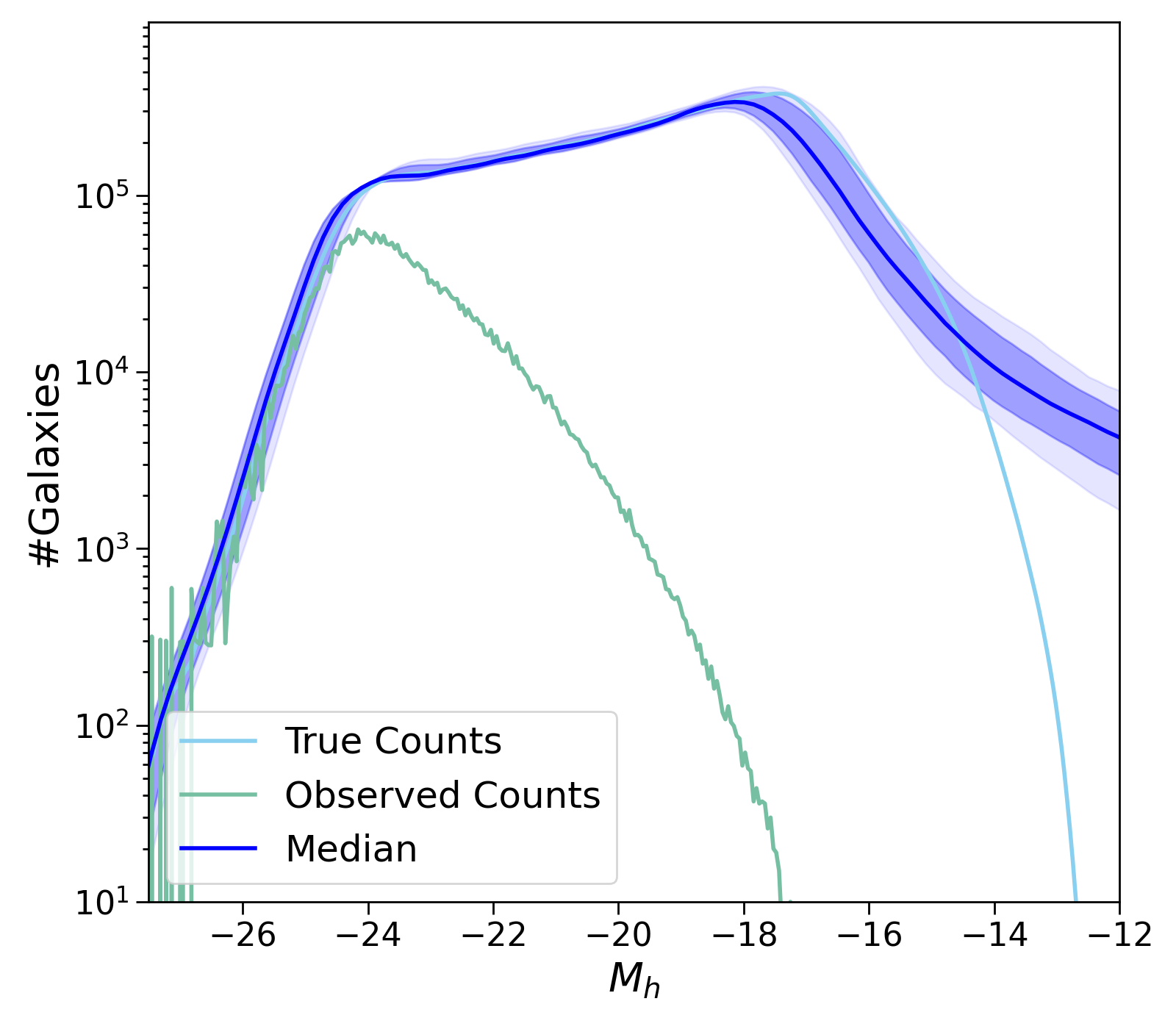}
    \caption{
    A histogram in the absolute magnitude $\Mhnorm$ (cf.~Eq.~\eqref{eq: def Mhnorm}). 
    The light blue (green) histogram includes all (detected) galaxies.  
    While the two distributions are almost identical at the bright end (low $\Mhnorm$), the detected galaxy number quickly plummets for fainter galaxies.
    Also note that the finite resolution of the Millennium simulation and its respective limited synthetic galaxy catalog cause an unphysical fall-off at $\sim -17$, visible in the true absolute magnitude distribution.
    The reconstruction and its associated $1\%$, $10\%$, $90\%$ and $99\%$ percentiles are drawn in dark blue.
    The true absolute magnitude distribution is well recovered up to $\Mhnorm\approx-18$.
    }
    \label{fig:reconstructed magnitude distribution}
\end{figure}
Fig.~\ref{fig:reconstructed magnitude distribution} shows the absolute magnitude distribution of all (detected) galaxies in light blue (green), and the reconstructed galaxy Poisson rate in dark blue.
The reconstruction recovers the full distribution faithfully, up to faint galaxies \mightchange{($\Mhnorm\sim -18$)}, where $\Mhnorm$ is defined in Eq.~\eqref{eq: def Mhnorm}. 
Since we reconstruct the galaxy rate only up to magnitudes above this value, the deviation between the true distribution and our reconstruction does not impact the final estimate of the number counts, as discussed in Sec.~\ref{subsec:impact Mthreshold}. 
The faithful recovery of the true magnitude distribution also justifies the generic galaxy magnitude modeling through a correlated Gaussian field that was introduced in App.~\ref{app:magnitude_distribution}. It is universal enough to capture the non-trivial galaxy magnitude distribution of \cite{Bertone:2007sj}.

\subsection{Impact of fixing the Hubble constant}
\label{subsec: impact fixing H0}

Since the ultimate motivation of this work is to infer the Hubble constant with GWs and improved redshift information from ``completed'' galaxy catalogs, it is crucial to assess the impact of the $H_0$ prior on the reconstruction.
To this purpose, we vary the analysis of the previous section with three different assumptions on the prior of $H_0$. 
(1) Fixing $h=0.67$ to its true value, with $h\deffrom H_0 /  (100 \hu)$. 
(2) A uniform prior \mightchange{$\p(h) = \mathcal{U}(0.6,0.8)$ (the result of Sec.~\ref{sec:result vanilla}).
(3) Fixing $h=0.73$, differing from its true value. }

\begin{figure}[ht]
    \centering
    \includegraphics[width=0.9\linewidth]{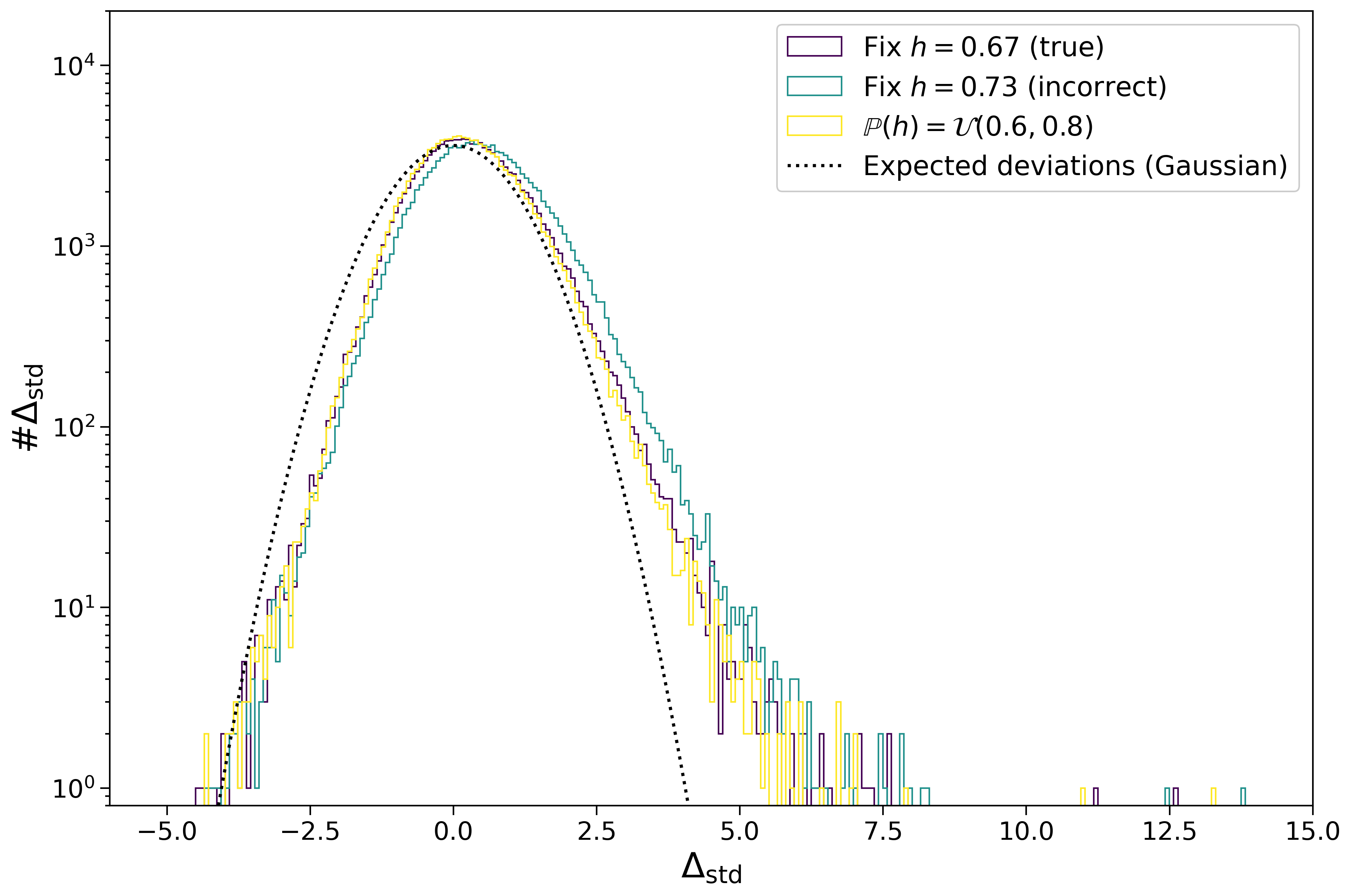}
    \caption{The histogram of the relative deviations, as a function of different $H_0$ values assumed. 
    }
    \label{fig: relative deviations histogram H0 impact}
\end{figure}

Fig.~\ref{fig: relative deviations histogram H0 impact} shows the comparison of the histogram of the relative deviations $\Deltastd$ (cf.~Eq.~\eqref{eq:def Delta std}), in the three aforementioned cases. 
The two distributions (fixing  $h=0.67$ and estimating $h$) are consistent, indicating that fixing $H_0$ to its correct value does not strongly impact the reconstruction. However, incorrectly assuming \mightchange{$h=0.73$} causes an underestimation of the uncertainties. 
The conclusion is thus to jointly estimate $H_0$ with the reconstruction of the galaxy Poisson rate rather than fixing it to a potentially incorrect value. 

\begin{figure}[ht]
    \centering
    \includegraphics[width=\linewidth]{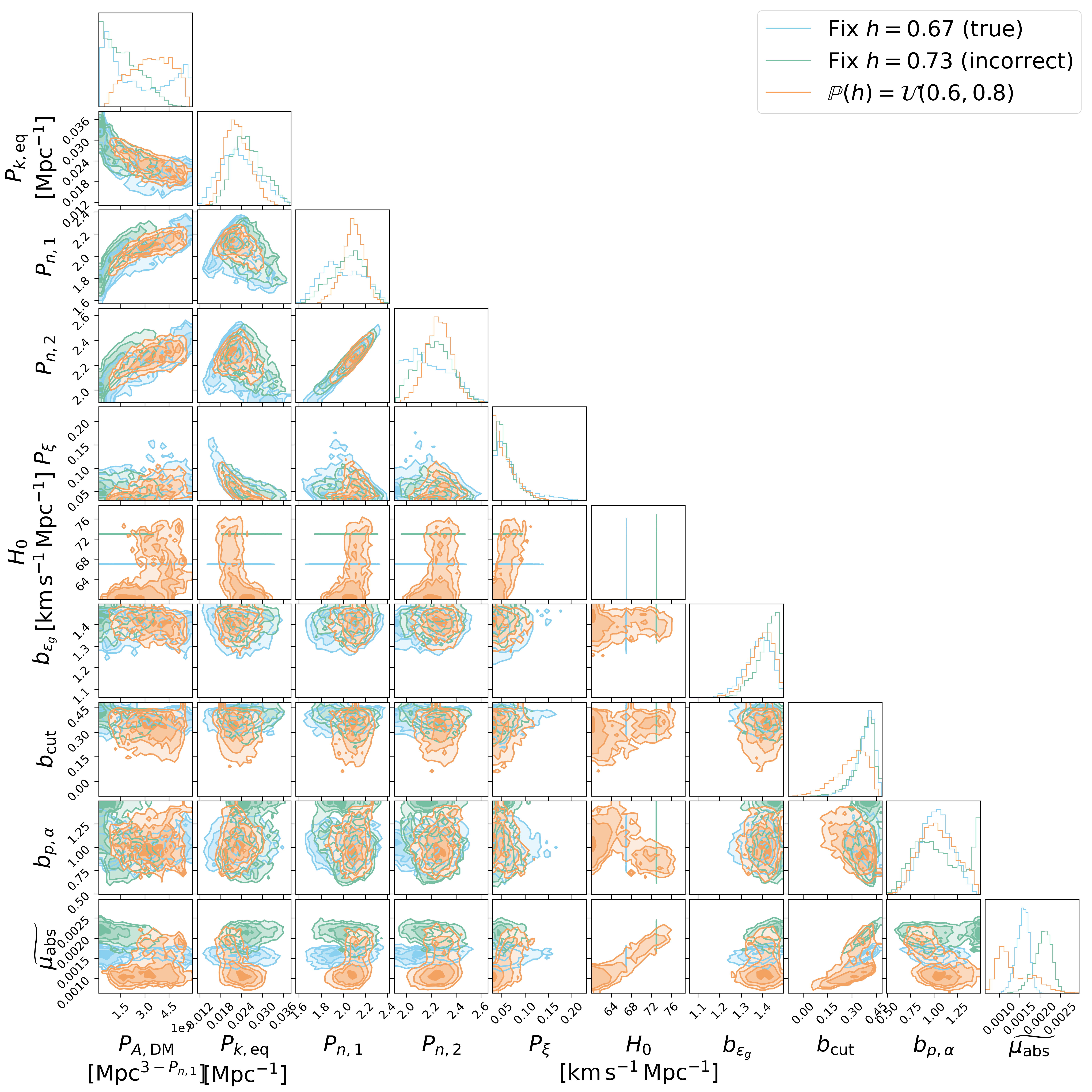}
    \caption{Corner plot for the inferred DM power spectrum parameters (cf.~Eq.~\eqref{eq: def phenomenological power spectrum}), the Hubble constant $H_0$, the bias parameters (cf.~Eq.~\eqref{eq:def galaxy bias}) and the overall rate (cf.~Eq.~\eqref{eq:def rate modified}) as a function of the $H_0$ prior. The behavior for the one-dimensional $H_0$ posterior in two cases is due to it being fixed.
    We plot the 0.5, 1, 1.5 and 2~2D-sigma contours. 
    }
    \label{fig:corner plot}
\end{figure}

For reference, Fig.~\ref{fig:corner plot} shows the corner plot of all one-dimensional hyperparameters, $\hyperparams$, i.e.~excluding the DM density contrast, the magnitude model and the detection probability (since the corner plot would be too large to include $\mathcal{O}(10^6)$ parameters). 
Let us stress that the $H_0$ posterior is dependent on the prior of the rate $\rateAbs$, since both parameters determine the overall number count. The impact of the rate prior on the $H_0$ measurement has thus to be assessed carefully.

\subsection{Impact of redshift uncertainties}
\label{subsec: impact redshift uncertainties}

\begin{figure}[ht]
    \centering
    \includegraphics[width=1.0\linewidth]{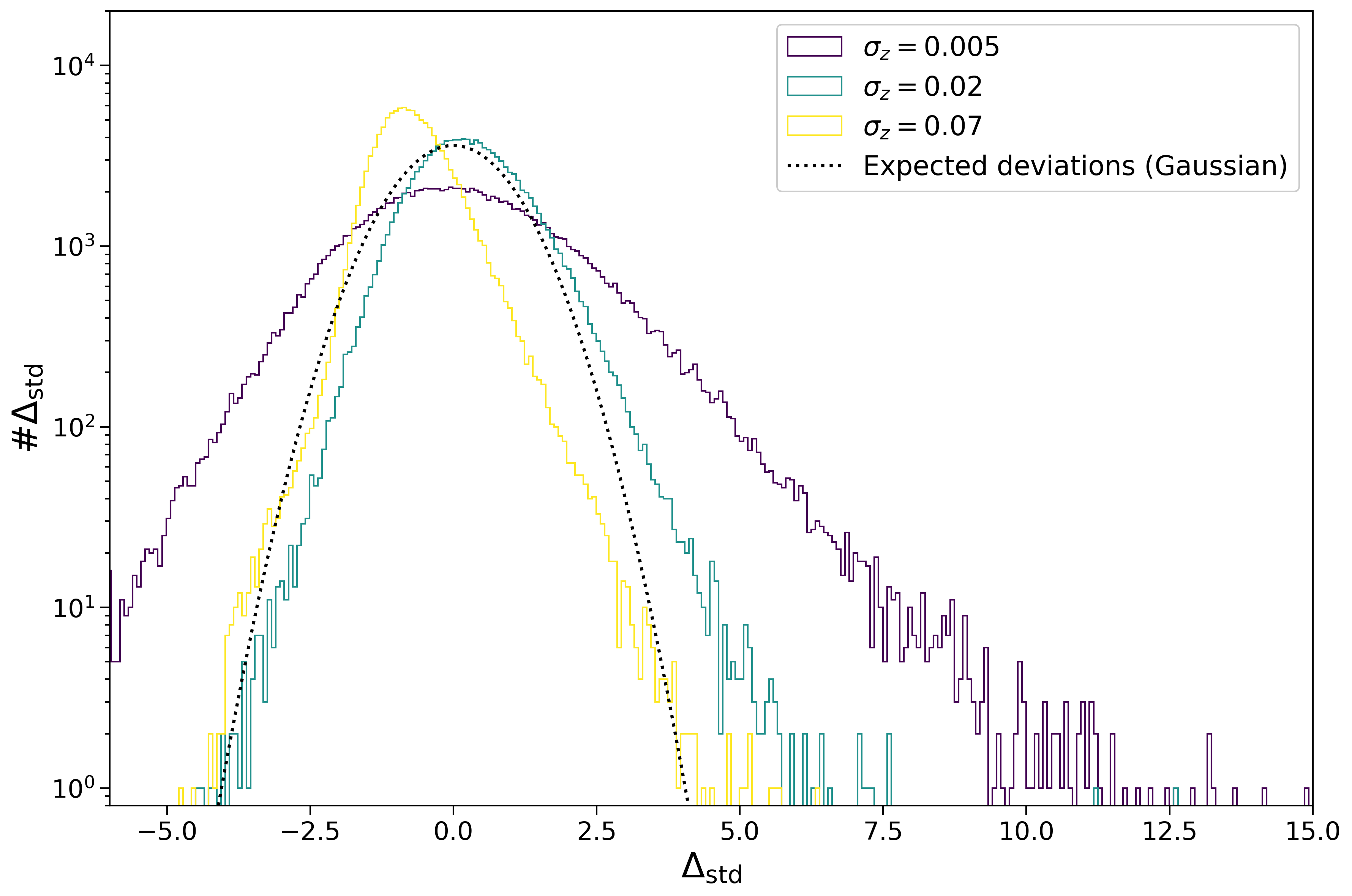}
    \caption{The deviations, $\Deltastd$ (cf.~Eq.~\eqref{eq:def Delta std}), as a function of the assumed relative uncertainties in the redshift. }
    \label{fig:impact redshift uncertainties}
\end{figure}

For the simulation of the observed galaxy number counts, we assumed a Gaussian distribution for the redshift uncertainties, with an evolving standard deviation of $\sigma_z = 0.01\times (1 + z)$.
This is a reasonable assumption for upcoming spectroscopic surveys \cite{DESI:2013agm, DESI:2016fyo, LSSTScience:2009jmu}, but it is important to understand how the results depend on the modeled redshift uncertainty as we assume it is constant.
To this end, we repeat the analysis of Sec.~\ref{sec:result vanilla} for different values of $\sigma_z$ (fixing $H_0 = 67\,\hu$).
We assume (\textit{i}) $\sigma_z = 0.005$, (\textit{ii}) $\sigma_z = 0.02$ and (\textit{iii}) $\sigma_z = \mightchange{0.07}$, using the same simulated redshift uncertainties, i.e.~$\sigma_z=0.01\times (1 + z)$ in all cases.
In Fig.~\ref{fig:impact redshift uncertainties}, we show the deviations $\Deltastd$ (cf.~Eq.~\eqref{eq:def Delta std}) for each of these three analyses.
As explained in a previous \mightchange{footnote}, the $\Deltastd$ distribution should follow approximately a zero-mean, unit-variance normal distribution. 
In the case of $\sigma_z = \mightchange{0.005}$ (\mightchange{violet}), the variance of the deviations is too large, indicating that the galaxy rate uncertainties are underestimated, as expected: the redshift uncertainties are too small when compared to the simulated uncertainties.
In the case of $\sigma_z = \mightchange{0.02}$ (\mightchange{blue}), the deviations are significantly closer to the zero-mean unit-variance Gaussian distribution. 
Finally, in the case of $\sigma_z = \mightchange{0.07}$ (\mightchange{yellow}), the deviations show a variance that is too small -- the uncertainties are overestimated.

We thus conclude that in order to be conservative, one should either (\textit{i}) account for the redshift-dependent uncertainties or, (\textit{ii}) use a larger redshift uncertainty overall that the effectively accounts for the redshift-evolving uncertainties.
While the first option is the most accurate, it is also the most computationally expensive as the process to compute $\ratezhathpm$ will no longer be a convolution with a Gaussian kernel (cf.~Eq.~\eqref{eq:redshift uncertainty rate}). \revvone{Depending on the rate of change of the width of the convolution with redshift, the computational cost can increase by an order of magnitude or more. 
Let us note here that the proposed framework is also suitable to analyze catalogs with both photometric and spectroscopic redshifts but requires a modified likelihood that accounts for the different redshift uncertainties. This analysis would also entail inferring two different selection effects, one for each measurement type. 
}

\subsection{Impact of the absolute magnitude threshold}
\label{subsec:impact Mthreshold}

\begin{figure}[ht]
    \centering
    \includegraphics[width=1\linewidth]{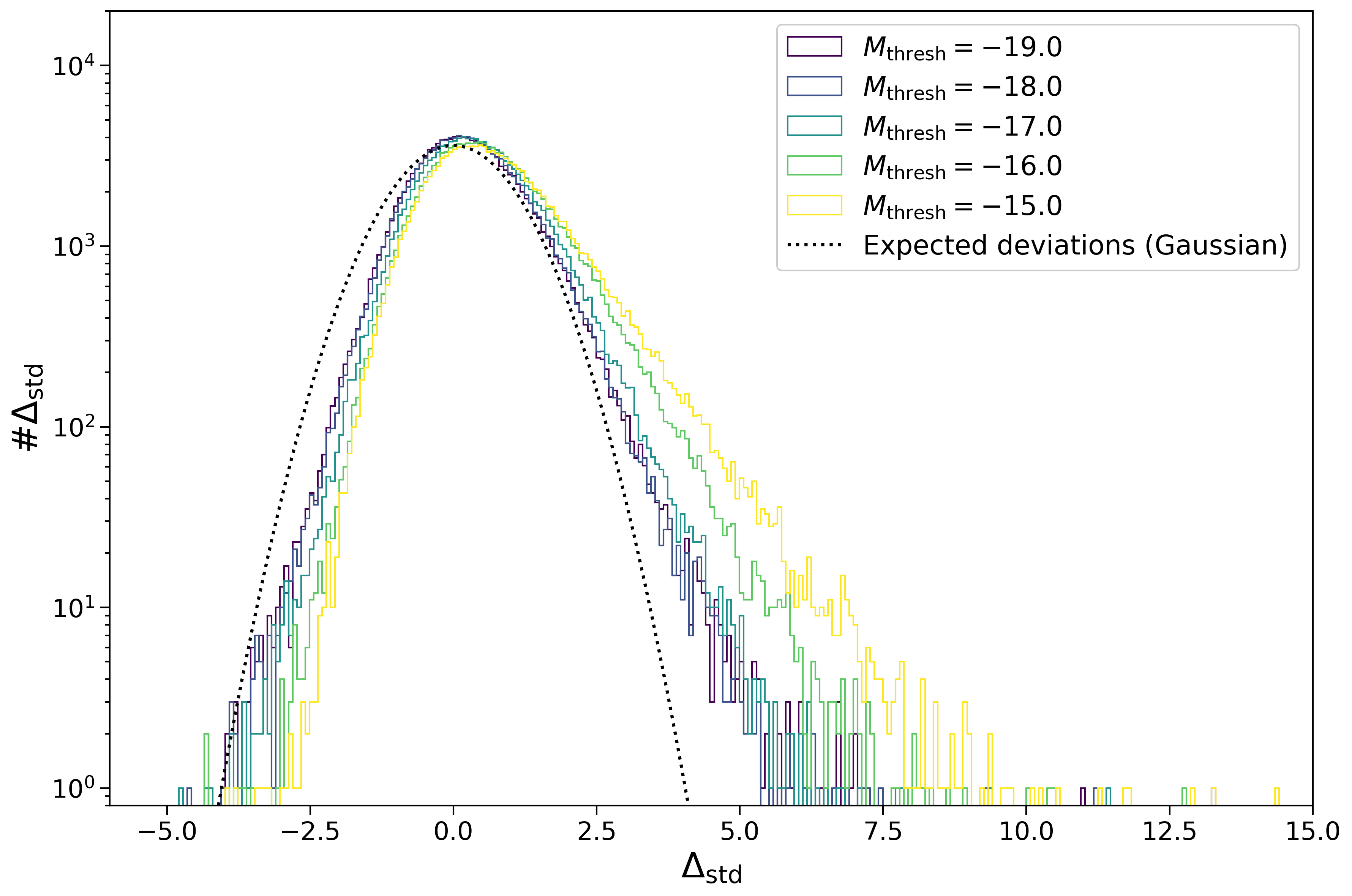}
    \caption{Histogram of the deviations, $\Deltastd$, defined in Eq.~\eqref{eq:def Delta std}. We also include the expected histogram if the posterior was Gaussian (dashed line). 
    As we move to include fainter galaxies (increasing the absolute magnitude threshold), the $\Deltastd$ histogram deviates increasingly from the theoretically expected histogram.
    }
    \label{fig:deviations std varying with M threshold}
\end{figure}

\begin{figure}[ht]
    \centering
    \includegraphics[width=0.8\linewidth]{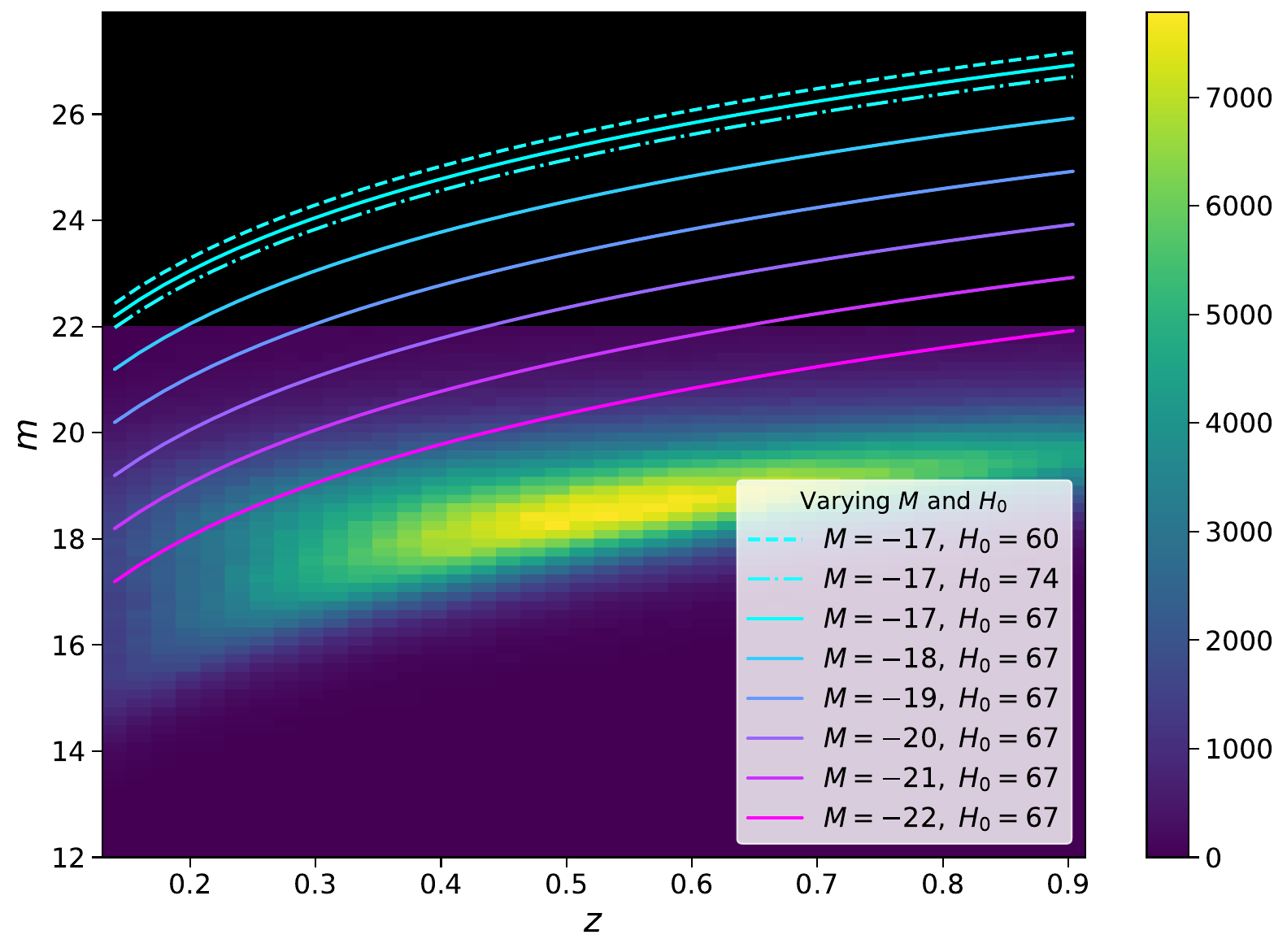}
    \caption{The apparent magnitude-redshift observed galaxy number counts, with lines corresponding to a given absolute magnitude. We also vary the assumed $H_0$ in order to show the resulting shift in apparent magnitude.
    Points outside the histogram that is used for the reconstruction are plotted in black. 
    }
    \label{fig: illustration apparent magnitude}
\end{figure}

Our inferred absolute magnitude distribution is faithful only up to a given threshold, determined by the faintest near-by galaxies.\footnote{Note that this assumes that the absolute magnitude distribution only weakly varies with redshift. }
Hence, when we infer the galaxy Poisson rate in each four-dimensional redshift, sky position and absolute magnitude bin, we set all counts above an absolute magnitude threshold, denoted as $\Mthresh$, to zero.
Ideally, this threshold coincides with the absolute magnitude of the faintest galaxy that has a non-negligible GW host probability.\footnote{In practice, this is difficult, as the galaxy GW host probability is not understood.}
If the probability of hosting a GW event is negligible for all galaxies above some $\Mthresh$ (as is expected from scaling relations relating star formation rates to the overall galaxy luminosity \cite{Artale:2019tfl}), we can safely ignore all galaxies with $M\geq\Mthresh$. 
This section varies this threshold $\Mthresh$ and studies its impact on the posterior of the galaxy Poisson rate.

We begin by varying the absolute magnitude threshold $\Mthresh$ for the result of Sec.~\ref{sec:result vanilla} in Fig.~\ref{fig:deviations std varying with M threshold}.
The histogram shows the deviations $\Deltastd$ (cf.~Eq.~\eqref{eq:def Delta std}) and compares their distribution to the expected distribution for a Gaussian posterior.
There is a clear trend that as we include fainter galaxies (i.e.~increasing $\Mthresh$), the distribution of $\Deltastd$ deviates increasingly from the Gaussian distribution. These deviations remain relatively small, especially when compared to other effects as varying the Hubble constant (cf.~Fig.~\ref{fig: relative deviations histogram H0 impact}). 
This is a consequence of the constraining power of the data -- the inferred absolute magnitude distribution deviates from the true distribution of absolute magnitudes more strongly for large values of the absolute magnitude, as illustrated in Fig.~\ref{fig:reconstructed magnitude distribution}.
The absolute magnitude above which the data provides no constraining power can be determined from Fig.~\ref{fig: illustration apparent magnitude}.
We vary here the absolute magnitude and show its corresponding apparent magnitude for a given redshift.
For low redshift, we see that the data allows for the observation of the faintest galaxies with an absolute magnitude of \mightchange{$M\sim -17$}. 
This is consistent with the aforementioned Fig.~\ref{fig:deviations std varying with M threshold}, where we see that the deviations started to significantly differ from the Gaussian distribution for \mightchange{$\Mthresh \gtrsim -17$.}


\section{Conclusions}
\label{sec: conclusions}

Gravitational waves from \glspl{cbc} are a promising avenue to measure the cosmic expansion history independently of traditional methods that solely use electromagnetic information. 
To constrain the cosmological parameters with GWs, the main bottleneck is obtaining accurate redshift information of the \gls{cbc}.
The dark siren method addresses this. It leverages galaxy catalogs into a redshift measurement by ``overlaying'' the GW sky position with the catalog.
As such, the galaxy catalog provides a redshift-sky position prior for the GW events. 
However, current methods that combine the GW data with galaxy catalogs only slightly improve the $H_0$ precision when compared to using GW data alone. This is mostly due to the limitations of the employed catalogs, which only contain galaxies up to $z\sim 0.3$ -- galaxies above that redshift are too faint to be confidently observed.
To address this, this work exploits the spatial correlations (through large-scale structure) -- not used in the standard codes -- to reconstruct the full galaxy density field. 
Our reconstruction improves the redshift–sky position prior for the GW signal, and through the more accurate redshift prior for GW events will lead to more informative constraints on the Hubble constant, something we plan to demonstrate in the near future.

We have demonstrated a Bayesian approach to reconstruct a galaxy catalog from a magnitude limited survey. 
To this end, we have simultaneously reconstructed the galaxy magnitude distribution, the DM power spectrum, the DM density contrast, the DM-galaxy bias, the Hubble constant and selection effects.
%
All the above ingredients are indispensable to connect the observed catalog to the full galaxy field.
We have demonstrated that one can recover the original (true) catalog to a high degree, and that the inferred uncertainties in the galaxy rate are accurate. 
Through our agnostic Gaussian process magnitude distribution, our method does not require a priori information about the galaxy absolute magnitude distribution or its parametric form. \revvone{This is with the caveat that this distribution was assumed to be independent of the DM density contrast --the effect of high density clusters producing brighter galaxies was neglected in this approach. }

The estimation of the magnitude distribution also allows for the inference of the galaxy Poisson rate in each magnitude bin, which is of importance since the connection between galaxy properties and GW host probability is not well understood. 
From one single reconstructed catalog we can thus test different weighting schemes: varying the link between the absolute galaxy magnitude and the galaxy GW host probability, allowing one to explore the impact of this systematic. 

We want to stress that this work prepares for a change in methodology for joint analyses of GW and galaxy catalog data. 
Ideally, a hierarchical approach integrates both data types (denoted as $\Data_{\rm EM }$ and $ \Data_{\rm GW }$, respectively) on equal footing. 
This approach avoids the preferential treatment of one data type over the other, a current limitation in the standard codes that treat the electromagnetic data as prior information, only evaluating the GW likelihood. In more mathematical terms, we advocate for a method that evaluates the likelihood $\p(\Data_{\rm EM }, \Data_{\rm GW }|\hyperparams)$ rather than $\p(\Data_{\rm GW }|\hyperparams, \Data_{\rm EM })$, as implemented at present. 
This will ensure that the cosmological parameter constraints are as robust and as accurate as possible. 

While we go significantly beyond the approximations that exist in the literature of current completion techniques in GW cosmology, it is important to acknowledge the simplifications made in our analysis. 
For instance, we assumed Gaussian errors for redshift measurements and did not account for possible uncertainties in apparent magnitude.
Additionally, our model supposes that the DM power spectrum is redshift-independent, i.e.~the amplitude of our structure is not growing. 
Future work will thus need to develop a more sophisticated model for the dark matter distribution; in addition to the overall amplitude redshift evolution also accounting for the nonlinear evolution at small scales. 
Furthermore, we have approximate the true galaxy number count by the galaxy Poisson rate since obtaining posterior samples in discrete number counts is computationally challenging.
Despite these simplifications, our method yields accurate line-of-sight redshift priors, which will enhance the precision of $H_0$ measurements.

Our work is not limited to the application of ``completed'' galaxy catalogs to GW data. 
More generally, our approach highlights that a hierarchical parameter estimation approach can be used to improve the parameter estimation of single galaxies, i.e.~the knowledge of the properties of neighboring galaxies can inform the properties of individual galaxies.

The present work, by simultaneously reconstructing cosmological and astrophysical parameters \`a la Bayes, offers a powerful framework for addressing limitations of the GW dark siren method. 
This advancement paves the way for improved $H_0$ constraints from GW and galaxy catalog data.

\section*{Acknowledgments}
We thank Sesh Nadathur, Rachel Gray, Simone Mastrogiovanni, Charles Dalang and Michael Williams for helpful discussion. K.L. and T.B. are supported by ERC Starting Grant SHADE (grant no.\,StG 949572). T.B. is further supported by a Royal Society University Research Fellowship (grant no.\,URF$\backslash$R$\backslash$231006).
Numerical computations were carried out on the \texttt{Sciama} High Performance Computing (HPC) cluster, which is supported by the Institute of Cosmology and Gravitation (ICG), the South-East Physics Network (SEPNet) and the University of Portsmouth.

\appendix

\section{Gaussian distributions}

\subsection{Gaussian Magnitude Distribution}
\label{app:magnitude_distribution}

In this appendix, we describe the mathematical formulation of our model for the absolute magnitude distribution. This approach is based on generating a gaussian random field in one dimension and modulating it with a power spectrum to introduce correlations. This approach is exactly analogous to the generation of the logarithm of the DM density contrast, where we are restricted to one dimension for the magnitude).
During inference, we determine the whitened parameters of the field, which represents a simpler sampling problem since the parameters are less correlated (as in the case of the whitened parameters of the DM density contrast (cf.~Sec.~\ref{subsec:DM model})). 
Note that the magnitude distribution is defined in the variable $\Mhnorm$, that we construct via
\begin{equation}
\label{eq: def Mhnorm}
    \Mhnorm \deffrom 
    M
    - 5 \log_{10}\left(\frac{H_0}{H_{0,{\rm ref}}} \right)
    \,,
\end{equation}
with $H_{0,{\rm ref}}$ a reference value of $H_0$, at which we have $\Mhnorm=M$. 
This definition ensures that the inferred distribution is independent of the assumed value of the Hubble constant. (Of course, if we want to reconstruct the distribution $\p(M|\hypermagnitudes)$, we will need to introduce a fiducial $H_0$ value).

\begin{figure}[ht]
    \centering
    \includegraphics[width=0.8\textwidth]{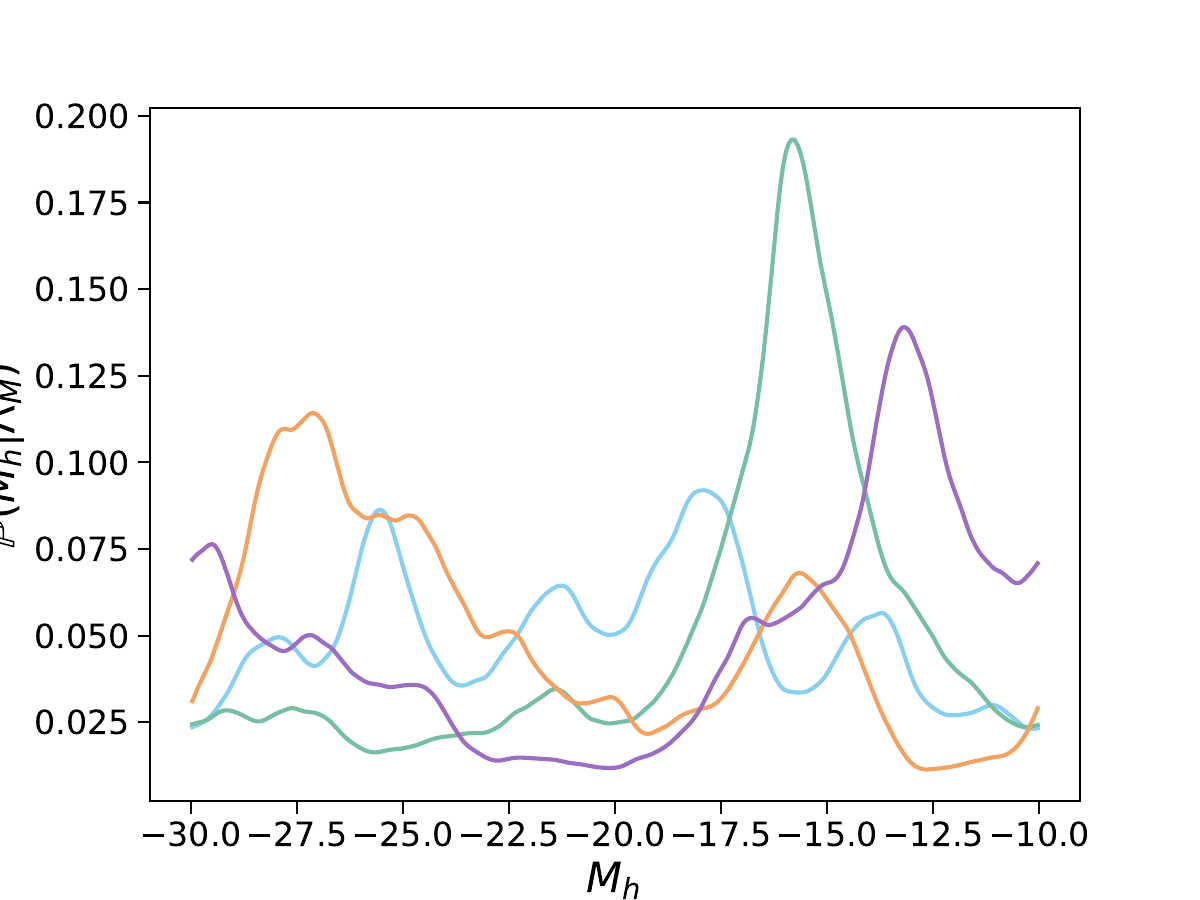}
    \caption{Visualization of the gaussian magnitude distribution. The structure of the field is determined by the chosen power spectrum, allowing for flexible modeling of magnitude distribution.}
    \label{fig: gaussian_magnitude_distribution}
\end{figure}

To construct the magnitude distribution, we define a two-point function that governs the correlation of the field. The specific form we adopt is a one-dimensional evolving power-law power spectrum, given by:

\begin{equation}
\label{eq:choice phenomenological power spectrum}
    P(k) = \MPSamp \cdot k_{\text{eff}}^{\Mpalph + \Malphs \log(k_{\text{eff}} / \Mkz)}
    \,,
\end{equation}
where
\begin{equation}
    k_{\text{eff}} = \sqrt{k^2 + \text{cut-off}}
    \,,
\end{equation}
and we assume a small regularization value \( \text{cut-off} = 10^{-6} \). In this expression:

\begin{itemize}
    \item \( \MPSamp \) controls the amplitude of the power spectrum,
    \item \( \Mpalph \) determines the power-law exponent,
    \item \( \Malphs \) introduces a running of the power-law exponent,
    \item \( \Mkz \) sets the (inverse of the) characteristic scale of the $k$-dependence of the power-law exponent.
\end{itemize}
\noindent
Throughout, we fix the parameters to \( \MPSamp = 0.06 \), \( \Mpalph = -1.5 \), \( \Malphs = -0.8 \), and \( \Mkz = 0.2 \). 
These particular values with the power spectrum structure allow a wide range of the possible magnitude distributions. While the specific parameter values above represent one possible choice, as demonstrated in the main text and in Figure~\ref{fig: gaussian_magnitude_distribution}, this framework is sufficiently flexible to recover the Millennium magnitude distribution, which has no particularly regular shape, while also enforcing a smooth behavior of $\p(\Mhnorm|\hypermagnitudes)$.
We have also verified that changing the parameter values by a factor of 2 does not impact the final result. 

A visualization of the variety of different random draws of the gaussian magnitude distribution, illustrating its structure and key properties, is provided in Figure \ref{fig: gaussian_magnitude_distribution}.

\subsection{Detection probability}
\label{app: cumulative gaussian random field model for detection probability}

\begin{figure}[ht]
    \centering
    \includegraphics[width=0.8\textwidth]{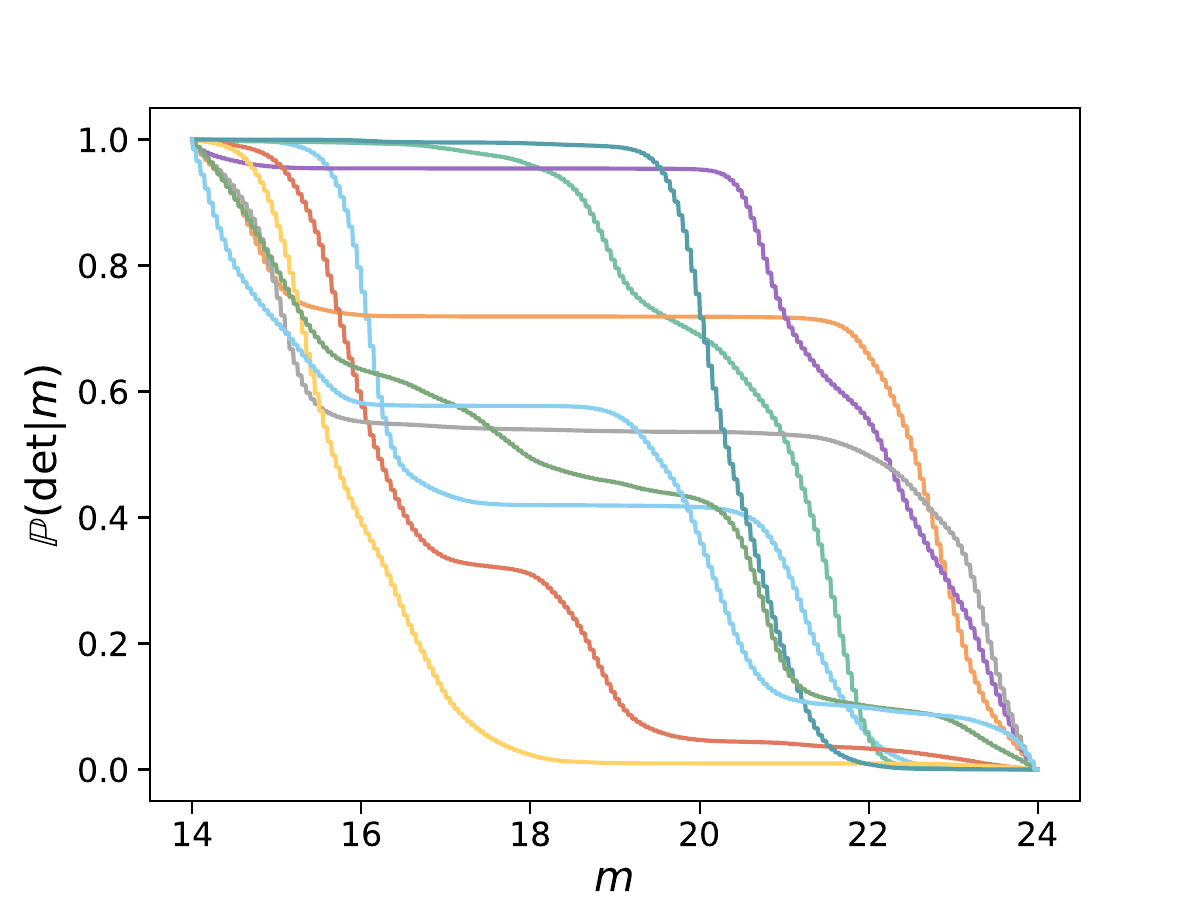}
    \caption{Visualization of the gaussian detection probability modeling. The structure of the variation is determined by the chosen ``power spectrum``, allowing for flexible modeling of detection probability.}
    \label{fig: detection_probability_visualization}
\end{figure}

Following the exact same mathematical modeling as for the flexible magnitude distribution, we use a one-dimensional Gaussian random field with the phenomenological power spectrum featuring in Eq.~\eqref{eq:choice phenomenological power spectrum} to generate random draws in an unphysical variable, $\Xnorm$. The apparent magnitude and $\Xnorm$ can be related by 
\begin{equation}
\label{eq:def Xnorm}
    \Xnorm \deffrom \frac{m - \mmean}{\mscale} \,.
\end{equation}
To obtain a strictly decreasing detection probability for increasing apparent magnitude, we then define
\begin{equation}
    \p({\rm det}|m) = 
    \int_{\Xnorm(m\rightarrow \infty)}^{\Xnorm(m)} G_{1D}(\Xnorm) \,\dd \Xnorm\,,
\end{equation}
where $G_{1D}$ is a random draw of the one-dimensional Gaussian random field. 
Throughout this work, we fix $\mmean=19$ and $\mscale=-1$, as well as the power spectrum parameters for the field as $\MPSamp = 10, \Mpalph=-1, \Malphs=-0.6$ and $\Mkz=0.1$, using Eq.~\eqref{eq:choice phenomenological power spectrum}. 
Fig.~\ref{fig: detection_probability_visualization} justifies the choice of these parameters: it allows for a sufficiently smooth model, while simultaneously being flexible enough to capture a wide variety of possible functional forms. Varying these parameters does not strongly impact the final inferred detection probability (plotted in Fig.~\ref{fig:pdet estimate vanilla}).

Note that we discretize the detection probability of grid size 200, which spans the range $(\Xnorm - 5, \Xnorm + 5)$. Thus, inferring the detection probability adds an additional 200 free parameters that we infer.

\section{Interpolation from cartesian to spherical coordinates}
\label{appendix:spherical_interpolation}

\begin{figure}[ht]
    \centering
    \includegraphics[width=0.8\linewidth]{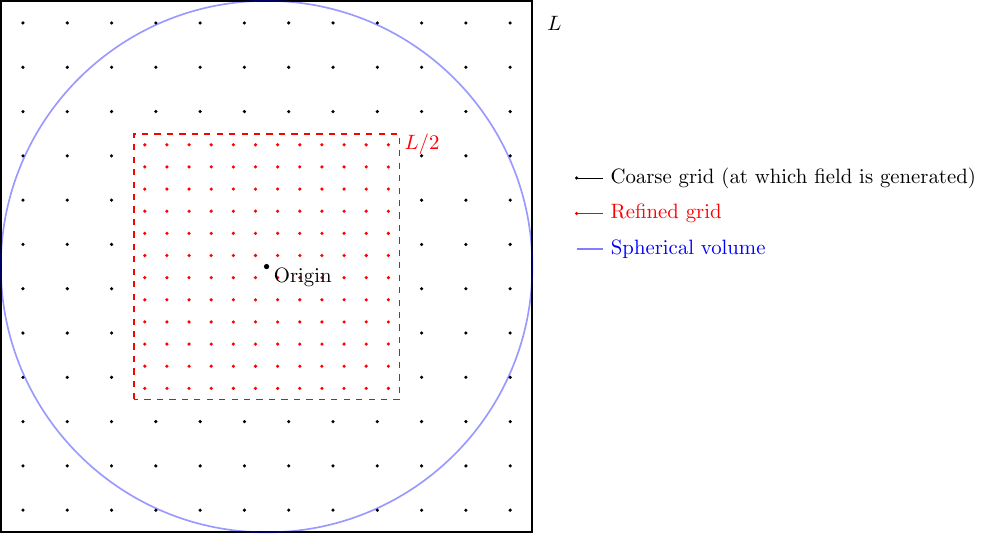}
    \caption{Two-dimensional slice to illustrate the spherical interpolation: the cartesian grid with refinement in the central region---for each coarse grid point (black, points not shown in the red region) we generate $8=2^3$ surrounding fine grid points (red region) that all share the same central value. For the reconstruction we typically use $\mathcal{O}(100) $ grid points per dimension. }
    \label{fig:refined cube schema}
\end{figure}

To be able to use the existing infrastructure of the log-normal field from the previous project \cite{Leyde:2024tov} based on \cite{Coles:1991if, Agrawal:2017khv}, this section summarizes how we connect the field in cartesian and spherical coordinates. 
For densities, the field value of the spherical element is then simply the average of the field values of the cartesian elements that fall inside that spherical element.
Therefore, we require that at least one center of a cartesian volume element to fall inside each of the three-dimensional spherical volume elements.
For a given spherical discretization this implies the minimum density of cartesian elements.
However, since the spherical elements are densely packed at the origin, this would require a very fine cartesian grid, with many cartesian elements in the region far from the origin.
Thus, we refine this method in the following way: centered at the origin, we define a cubic region with side length $L/2$ (for $L$ the length of the total cube) at which all cartesian elements are repeated twice along each dimension.
See Fig.~\ref{fig:refined cube schema} that illustrates the denser three-dimensional grid in the central region of the simulated cube.

\section{Prior range}
\label{app: prior range}

Table~\ref{tab: priors} lists our prior choice. 
Note that we choose to infer a modified rate parameter $\rateAbstilde$, rather than $\rateAbs$ that are related through
\begin{equation}
\label{eq:def rate modified}
    \rateAbstilde \deffrom 
     \frac{\rateAbs}{\int \dd z\,\dd \skypos \,\densityDM(z,\skypos)}
    \,,
\end{equation}
saving us the computation of the normalization step of Eq.~\eqref{eq:pdf galaxies}. 

\begin{table}
\centering
\renewcommand{\arraystretch}{1.3} 
\begin{tabular}{lll}
\hline
\textbf{Variable} & \textbf{Name} & \textbf{Prior} \\
\hline
\hline
\multicolumn{3}{c}{\textbf{Cosmological parameters}} \\
\hline
$H_0$ & Hubble constant & \text{Varying.}\\
$\Omega_m$ & Matter density & $\delta(0.3)$ \\
\hline
\multicolumn{3}{c}{\textbf{Dark matter power spectrum parameters}} \\
\hline
$\PA$ & Power spectrum amplitude & $\mathcal{LU}(10^5,6\times 10^7)$ \\
$\PkEq$ & Power spectrum scale break & $\mathcal{LU}(2\times10^{-3},0.04)$ \\
$\PnOne$ & Low-$k$ slope & $\mathcal{U}(1.5,2.5)$ \\
$\PnTwo$ & High-$k$ slope & $\mathcal{U}(1.9,3.8)$ \\
$\Pxi$ & Transition sharpness & $\mathcal{LU}(0.03,0.25)$ \\
\hline
\multicolumn{3}{c}{\textbf{Galaxy bias model parameters}} \\
\hline
$\biasamp$ & Bias amplitude & $\delta(1.0)$ \\
$\biasepsilon$ & Bias modulation & $\mathcal{U}(0.1,1.5)$ \\
$\biascut$ & Bias exponential cut-off & $\mathcal{U}(-0.2,0.5)$ \\
$\biasalpha$ & Bias power-law slope & $\mathcal{U}(0.4,1.5)$ \\
\hline
\multicolumn{3}{c}{\textbf{Rate parameter}} \\
\hline
$\rateAbstilde$ & Overall rate & 
$\mathcal{LU}(10^{-4},10^{-2})$ \\
\hline
\multicolumn{3}{c}{\textbf{Magnitude Model Parameters}} \\
\hline
$\MPSamp$ & Power spectrum amplitude & $\delta(0.06)$ \\
$\Malphs$ & Power spectrum slope & $\delta(-0.8)$ \\
$\Mkz$ & Power spectrum break scale & $\delta(0.2)$ \\
$\Mpalph$ & Power spectrum shape & $\delta(-1.5)$ \\
\hline
\end{tabular}
\caption{Summary of priors used in the analysis. $\mathcal{U}(a,b)$ denotes a uniform prior, $\mathcal{LU}(a,b)$ a log-uniform prior, and $\delta(c)$ a delta function fixed at value $c$. The power spectrum for dark matter follows a double power-law model (cf.~Eq.~\eqref{eq: def phenomenological power spectrum}). The bias model corresponds to a power-law with exponential cut-off, cf.~Eq.~\eqref{eq:def galaxy bias}), and the magnitude model uses a Gaussian magnitude field (cf.~App.~\ref{app:magnitude_distribution}).
}
\label{tab: priors}
\end{table}

\bibliographystyle{unsrt}
\bibliography{references}

\end{document}